\documentclass[preprint,showkeys,showpacs,preprintnumbers,amsmath,amssymb]{revtex4}

\usepackage{graphicx}

\usepackage{mathrsfs} % required for $\mathscr J$ ...

\usepackage{amssymb}
\usepackage{amsmath}

\usepackage{mydefns}

\newcommand{\grad} {{\nabla}}		% gradient operator
\newcommand{\gradh} {{\nabla_h}}	% horizontal gradient operator
\newcommand{\vh} {\ve v_h}		% horizontal velocity

 \newcommand{\win}{\varpi}

 \newcommand{\mwtn}[1] {{\widehat {#1}^{\sim\win}}}
 \newcommand{\widemwtn}[1] {{\widehat { \left(#1\right) }^{\sim\win}}}
 \newcommand{\vhmwtn} {{\widehat {\ve v}_{h}^{\sim\win} }} % mwt of hor velo

\newcommand{\mwt}[3] {{\widehat {#1}_{#2}^{\sim{#3}}}}
\newcommand{\widemwt}[3] {{\widehat { \left(#1\right)  }_{#2}^{\sim{#3}}}}

\newcommand{\mwtr}[2] {{{#1}^{\sim #2}}}
\newcommand{\lon} {\lambda}	% longitude
\newcommand{\lat} {\varphi}	% latitude

\newcommand{\Dlon}[1] {{\frac {\D #1} {\D \lon} }}
\newcommand{\Dlat}[1] {{\frac {\D #1} {\D \lat} }}

\newcommand{\w}{\omega}		% pressure velocity =dp/dt
\newcommand{\gp}{{\Phi}}		% geoptential

\newcommand{\wwt}{\widehat}

\newcommand{\lgsw}{{\rm\sim 0}}
\newcommand{\mesosw}{{\rm\sim 1}}
\newcommand{\subsw}{{\rm\sim 2}}

\newcommand{\ext}{{\ell}}
\newcommand{\sclt}{\widehat}		% scaling transform

\newcommand{\lgr}[1]{{#1}^{\lgsw}}
\newcommand{\mesor}[1]{{#1}^{\mesosw}}
\newcommand{\subr}[1]{{#1}^{\subsw}}

\newcommand{\swt}[3]{\wwt {#1}_{#2}^{{\rm\sim} #3}}
\newcommand{\swr}[2]{{#1}^{{\rm\sim} #2}}

\newcommand{\swtnw}[1]{\swt {#1} n \win}

\newcommand{\transfer} {{\Gamma}}		

\newcommand{\recept} {{\cal R}}		

 \newcommand{\nlcoef}{{\alpha_\ell}}	% nonlinear coefficient (alpha)
\newcommand{\Fr}{{F_r}}		% rotational internal Froude # 
\usepackage{color}
\newcommand{\trackchanges}[1] {{#1}}
\begin{document}

\title{Canonical transfer and multiscale energetics for primitive and quasi-geostrophic atmospheres}

\author{X. San Liang\footnote{URL: http://www.ncoads.org/}}

\email{san@pacific.harvard.edu}
\affiliation{
	% Center for Ocean-Atmosphere Dynamical Studies,
	School of Marine Sciences and School of Atmospheric Sciences,\\ 
	Nanjing Institute of Meteorology, Nanjing 210044, China\\
	\vskip 2cm 
	{Preprint submitted to Journal of the Atmospheric Sciences} }
		% (DOI:10.1175/JAS-D-16-0131.1)} }
	% Theodore Research Institute, Massachusetts}

\begin{abstract}
{
The past years have seen the success of a novel multiscale energetics
formalism in a variety of ocean and engineering fluid applications.
In a self-contained way, 
this study introduces it to the atmospheric dynamical diagnostics,
with important theoretical updates.
Multiscale energy equations are derived using a new analysis apparatus,
namely, multiscale window transform, 
with respect to both the primitive equation and quasi-geostrophic models.
A reconstruction of the ``atomic'' energy fluxes on the 
multiple scale windows allows for a natural and unique separation of the 
in-scale transports and cross-scale transfers
from the intertwined nonlinear processes. The resulting
energy transfers bear a Lie bracket form, reminiscent of the Poisson bracket 
in Hamiltonian mechanics; we hence would call them
``canonical''. A canonical transfer process is a mere redistribution of energy 
among scale windows, without generating or destroying energy as a whole. 
  % With the technique the multiscale atmospheric kinetic energy 
  % (KE) and available potential energy (APE) equations are derived. 
By classification, a multiscale energetic cycle
comprises of available potential energy (APE) transport, 
kinetic energy (KE) transport, pressure work,
buoyancy conversion, work done by external forcing and friction,
and the cross-scale canonical transfers of APE and KE which
correspond respectively to the baroclinic and barotropic 
instabilities, \trackchanges{among others}, in geophysical fluid dynamics.
A buoyancy conversion takes place in an individual window only,
bridging the two types of energy namely KE and APE; 
it does not involve any processes among different scale windows, 
and is hence basically not related to instabilities. 
This formalism is exemplified with a preliminary application to the
Madden-Julian Oscillation study.
}
\end{abstract}
%\pacs{}


\maketitle

%\tableofcontents

\section{Introduction}
Ever since Lorenz (1955)\cite{Lorenz} % (Lorenz, 1955, refer to the review of Tailleux, 2013) 
introduced the concept of available potential
energy (APE), and set up a two-scale formalism of energy equations 
using the Reynolds decomposition, energetic analysis has become a powerful
tool for diagnosing atmospheric and oceanic processes. 
Related studies include mean flow-wave interaction (e.g.,
\cite{Dickinson}, \cite{Boyd}, \cite{McWilliams99}, 
\cite{Fels_Lindzen}, \cite{Matsuno}), 	% (Dickinson, 1969)
upward propagation of planetary-scale disturbances 
(\cite{Charney}), 	% (Charney and Drazin, 1961)
ocean circulation energetics 
(\cite{Holland}, \cite{Haidvogel}),  % (Holland, 1978),
mean current-eddy interaction (\cite{Hoskins}),  % (Hoskins et al., 1983),
atmospheric blocking (\cite{Trenberth}, \cite{Fournier}, \cite{Luo}), 
			% (Trenberth, 1986; Fournier, 2001), 
Gulf Stream dynamics (\cite{Dewar}),  % (Dewar, 1989), 
	% global atmospheric enegetics 
   normal modal interaction  (\cite{Sheng90}), % (Sheng and Hayashi, 1990), 
regional cyclogenesis (\cite{Cai90}), % (Cai and Mak, 1990), 
convection and cabbeling (\cite{Ingersoll16}), % (Su et al., 2016),
and the most recent studies such as \cite{Cai07},   % Cai et al. (2007), 
\cite{Waterman}, 	% Waterman and Jayne (2011), 
\cite{Murakami11},	% Muramuki (2011), 
\cite{Hsu},		% Hsu et al. (2011),
\cite{Chen},		% Chen et al. (2014), 
\cite{Chapman},		% Chapman et al. (2015), 
to name a few. 
Meanwhile, 
	% Realizing that two scales might not be enough for many problems,
Saltzman\cite{Saltzman57}	% Saltzman (1957) 
cast the problem into the framework of Fourier analysis,
and obtained the energetics in the wavenumber domain, while
Kao\cite{Kao}	% Kao (1968) 
further extended it to the wavenumber-frequency space. 
Now both approaches have become standard in geophysical fluid
dynamics and other fluid-related fields; see, for example, 
\cite{Pedlosky},	% Pedlosky (1987),
\cite{Chorin},		% Chorin (1994), 
\cite{Pope},		% Pope (2003), 
etc.

Lorenz' energetics in bulk form, i.e., in the form of global
mean or integral, have clear physical interpretations (e.g., \cite{Pedlosky}). 
This global mean form, however, may be inappropriate for regional
diagnostics, as real atmospheric processes are localized in nature; 
in other words, they tend to be locally defined in space and time, 
and can be on the move. 
The Madden-Julian Oscillation (MJO) that we will take a brief look
at the end of this study is such an example; it is a progressive 
process that involves energy production and dissipation.
For this reason, it has been a continuing effort to relax the spatial
averaging/integration to have these processes faithfully represented.
A tradition started by Lorenz himself
is to collect the terms in divergence form and combine them
as one term representing the transport process, separate 
the term from the nonlinear interaction and take the residue as the 
energy transfer between the distinct scales (e.g., \cite{Harrison}).
	% (e.g., Harrison and Robinson, 1978). 
Now this has been a standard approach to multiscale energetic diagnostics
for fluid research,
particularly for turbulence research, 
where much effort has been devoted 
to engineering the so-obtained transfers (cf.~\cite{Pope}).
		% (cf.~Pope, 2003). 

While we know a transport process indeed bears a divergence form in the 
governing equations, the above transport-transfer separation is not 
unique. Multiple divergence forms exist that may yield quite different 
transfers. As argued by Holopainen (1978), the resulting
energy transfer in such an open system is quite ambiguous. 
This issue, which is actually much profound in fluid
dynamics, has long been discovered, but have
not received enough attention, except for a few studies 
such as \cite{Plumb83}. 
	% Plumb (1983).	
(The author was just aware of Berloff's discussion on the consistency of
eddy fluxes\cite{Berloff}, which also seems to be related to this problem.)

Another major issue in formulating multiscale energetics regards the
machinery for process decomposition by scale. 
Traditionally two methods, namely,
Reynolds' mean-eddy decomposition (MED) and Fourier transform, have been used.
The former is originally a statistical notion with respect to an ensemble
mean, but for practical reason the ensemble mean is usually replaced by time
mean, zonal mean, etc., making it a tool of scale decomposition.
Both these methods are global, in the sense that they do not retain the
local information. This is generally inappropriate for realistic
atmospheric processes such as instabilities, which are in nature highly
localized energy burst processes. In remedy, a practical approach that 
is commonly used is to do running time mean over a chosen duration of time. 
Indeed this gives the local information while retaining the simplicity 
of the Reynolds formalism. However, it does not solve the fundamental
problem that an energy burst process, among others, is by no means
stationary over any duration; any scale decomposition under such a hidden
assumption may result in spurious information, preventing one from making 
correct diagnoses.

An alternative approach to overcoming the difficulty is via filtering.
Filters have been widely used to separate processes involving different
scales.
But for energetics studies, it seems that a very fundamental issue 
has been completely ignored, that is, how energy (and any quadratic
properties) should be expressed in this framework. 
Currently the common practice is, for a two-scale decomposition, 
to first apply some filter to separate a field variable, 
say, $u$, into two parts, say, $u_L$ and $u_S$, which 
represent the large-scale and small-scale features, and then take
$u_L^2$ and $u_S^2$ (up to some factor) as the large-scale and 
small-scale energies. While this intuitively based and widely used 
technique may be of some use in real problem diagnostics, 
it is not physically relevant---one immediately sees the
inadequacy by noticing that $u^2 \ne u_L^2 + u_S^2$. 
In fact, multiscale energy is a concept in phase space, such as that
in Fourier power spectra; it is related to physical energy through
a theorem called Parseval relation. Attempting to evaluate multiscale 
energies with the filtered (low-pass, band-pass, etc.) or
reconstructed field variables is conceptually off track.
Actually this is a difficult problem, and has not been well
formulated until filter banks and wavelets are connected 
	(\cite{Strang}).
	% (Strang and Nguyen, 1997).
Besides, energy conservation requires that the resulting subspaces from
filtering must be orthogonal, as we will elaborate in the following 
section. This requirement, unfortunately, has been mostly ignored in
previous studies along this line.

The other line in this regard is with respect to 
Fourier transform (\cite{Saltzman57}, and the sequels), 
which does not have local information retained, either.
Coming to remedy is wavelet transform or, to be precise, 
orthonormal wavelet transform (OWT), as only with an orthogonal basis 
can the notion of energy in the physical sense be introduced. 
	% In this regard, Farge (1992) pioneers the study, 
In 2000, Fournier first introduced OWT into the study of
atmospheric energetics. 
This is a formalism with respect to space.
While opening a door to % a colorful of 
localized spectral structures, 
many processes such as transports are
not as easy to see as those in the Lorenz-type formalisms.
%
% and hard to interpret. In fact, the complexity is
% increased exponentially with the number of scale;
% we will see later in this study (cf.~Fig.~\ref{fig:schematic}) that how
% the processes complexified as the scale number increases from 2 to 3. 
%
On the other hand, the atmospheric and oceanic processes tend to occur on
a ranges of scales (e.g., MJO has a scale range of 30-60~days),
or {\it scale windows} as we will introduce in the following section, 
rather than on individual scales. 
For OWT, transform coefficients (hence multiscale energies) 
are defined discretely at different locations for different scale levels; 
there is no way to add them through a range of scales to make 
an expression of localized energy for that range. These issues, among
others, are yet to be addressed with these formalisms.

% In this sense, the traditional mean-eddy decomposition used in the 
% Lorenz formalism is still the preferred diagnostic tool.

% In the wavenumber space, also the same transport issue needs to be
% addressed. Can the transfer-transport separation be achieved through
% interaction analysis?

So, to relax the spatial averaging in a bulk energetics formalism incurs the 
issue of transport-transfer separation, while to improve the MED to have local
information retained requires a more sophisticated machinery of scale 
decomposition.
Can we put these two issues in the same framework and solve them 
in a unified approach? The answer is yes. The early attempts include 
the multiscale oceanic energetics studies by 
	Liang and Robinson (2005)\cite{LR05} (LR05 henceforth),
	% Liang and Robinson (2005; LR05 henceforth), 
based on multiscale window transform (MWT),
a functional analysis tool which was rigorized later 
(Liang and Anderson 2007\cite{LA07}, LA07 hereafter).
%
%
% and has been our endeavor 
% The two issues, machinery for scale decomposition and separation of 
% interscale transfer from the transport process, 
%
This formalism has been mostly overlooked, though it has been applied with
success to a variety of real ocean problems 
	  (e.g., \cite{LR04}, \cite{LR09}) 
	%  (e.g., Liang and Robinson, 2004; 2009) 
and engineering problems 
	(e.g., \cite{Liang_Wang}),
	%  (e.g., Liang and Wang, 2004),
partly because it has not been introduced for atmospheric studies, 
and has not been formulated in spherical coordinates. 
	% preventing it from being applied to global diagnoses. 
(As we will see soon, expressing the energetics in spherical
coordinates is by no means an easy task.)
This study is purported to address these issues, giving a comprehensive
and self-contained
introduction of the fundamentals and the progress since LR05.
A key point that distinguishes this study from the earlier effort
is that, in LR05, the transport-transfer separation was introduced
in a half-empirical way. With the nice properties of the MWT which was
formally established later on in LA07, we will see soon in the next section
that this actually can be put on a rigorous footing, and the resulting 
transfer bears a Lie bracket form, reminding us of the Poisson bracket in
Hamiltonian mechanics.
Besides, in this study we will extend the formalism to 
quasi-geostrophic flows, which must be derived in a different way.
Considering the traditional and recently renewed interests 
(e.g., \cite{Murakami11}) in multiscale atmospheric energetics diagnostics, 
and \trackchanges{considering}
 that a topic of much concern in turubulence research
is to engineer the resulting transfer, 
this rigorous study is rather timely.

In the following 
we first give a brief introduction of the concepts of scale window, 
multiscale window transform, and multiscale energy. 
In \S\ref{sect:separation}, we show that how the flux on a specific scale 
window can be rigorously derived, and how the energy transfer between two
scale windows can be obtained. We will see that the resulting transfer 
bears a form like Lie bracket, reminding one of the Poisson bracket 
in Hamiltonian dynamics. We then derive the evolution equations for the 
multiscale kinetic energy (KE) and available potential energy (APE) with
both a primitive atmospheric model (\S\ref{sect:atmos}) 
and a quasi-geostrophic model (\S\ref{sect:QG}). For completeness, 
a summary of the multiscale oceanic energetics, together with the needed
modification, is briefly presented (\S\ref{sect:ocean}); also included
is a brief review of some necessary horizontal treatment 
(\S\ref{sect:interact_analysis}). 
In section~\ref{sect:example}, we demonstrate how the formalism may be 
applied, using the Madden-Julian Oscillation as an example. 
This study is summarized in \S\ref{sect:summary}.
For easy reference, in appendix~A a glossary of symbols is provided.
The related software can be downloaded
from the website http://www.ncoads.org/ 
(within the section ``Software'').
%
% \footnote
%    {
%    Just in case of any change, please search for the author's website.
%    }

\section{Multiscale window transform}
% * refute the modus operandi with traditional filters
% * mention why choose MWT in time

% As mentioned above, filtering is a good method to separate a signal
% by scale, while retaining the locality in physical space. However, 
% when energy (or any quadratic properties) is considered, a problem 
% arises in representing the energy on a designated scale. The common
% practice, which is simply a square of the reconstructed field on that
% scale, is conceptually incorrect and physically meaningless. In fact, 
% the energy representation is by no means an easy problem; only
% after filter banks were connected to wavelets was the representation 
% for the first time made possible.
% One may then wonder why the classical mean-eddy decomposition (MED) 
% through Reynolds averaging works well, with simple expressions for
% the two-scale energies. Toward the end of this section, 
% we will see that, the Reynolds averaing, though simple in form,
% actually has a profound meaning in functional analysis in that it 
% is both a transform and a reconstruction. We need to generalize it 
% to have local physics retained and, when necessary, to introduce more scales.
%
% The {\it multiscale window transform} developed by Liang and Anderson
% (2007; LA07 hereafter) is such a generalization. 
% This section gives it a brief introduction.

% Locality in physical space is kept and it allows for interactions 
% beyond the mean and eddy processes, e.g., the mean-eddy-turbulence 
% interaction. This section gives it a brief introduction.

\trackchanges{
This section gives a very brief introduction of the multiscale window transform
developed by {LA07}. The first part (\S\ref{sect:mwt}) is the fundamentals;
but the reader may simply skip it if he/she already knows the notation and the
fact that a reconstruction is conceptually different from a transform.
}

\subsection{Scale window and multiscale window transform}  \label{sect:mwt}

% An MWT organizes a signal into several distinct time scale ranges, 
% while retaining its track in physical space. 

More often than not, an atmospheric process tends to occur on a range of
scales, such as the MJO which has a broadband
spectrum between 30 and60 days (cf.~section~\ref{sect:example}),
rather than on individual scales. 
Such a scale range is called, in a loose sense, a
{\it scale window}. Rigorously it can be defined  over a univariate 
interval or a multi-dimensional domain. In this study, the former is used,
as we only deal with time. This is in accordance with Lorenz' formalism.
Historically it has long been discussed 
		 (e.g., \cite{Haynes}), 
		% (e.g., Haynes, 1988), 
and has been justified by the observational fact that, in the atmosphere, 
scales in time and in space are correlated. 
Besides, only scales defined over a univariate field can be unambiguously 
referred to as large scale, small scale, and so forth, 
as desired in the atmospheric energetics studies. 

\begin{figure*}[t]
   \begin{center}
   \includegraphics[angle=0,width=0.5\textwidth] {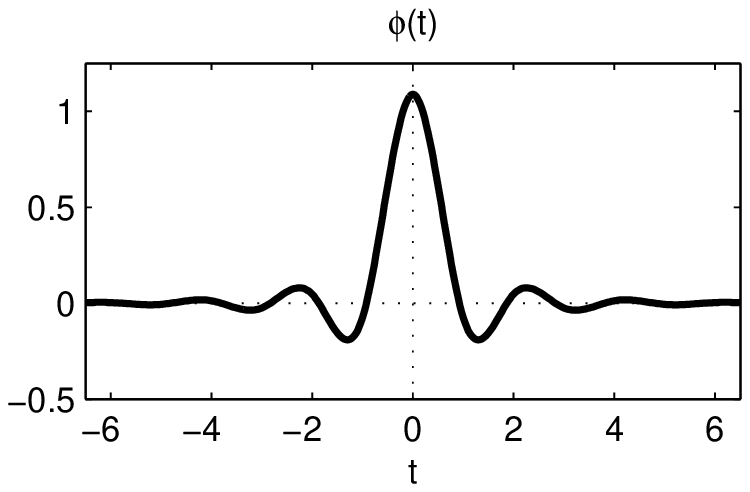}
   \caption
         {The orthonormal scaling function $\phi$ 
	  constructed in \cite{LA07}.
%	  via cubic spline orthonormalization 
         \protect{\label{fig:scl_func}}}
   \end{center}
   \end{figure*}

\begin{figure*} [t]
   \begin{center}
   \includegraphics[angle=0,width=0.9\textwidth] {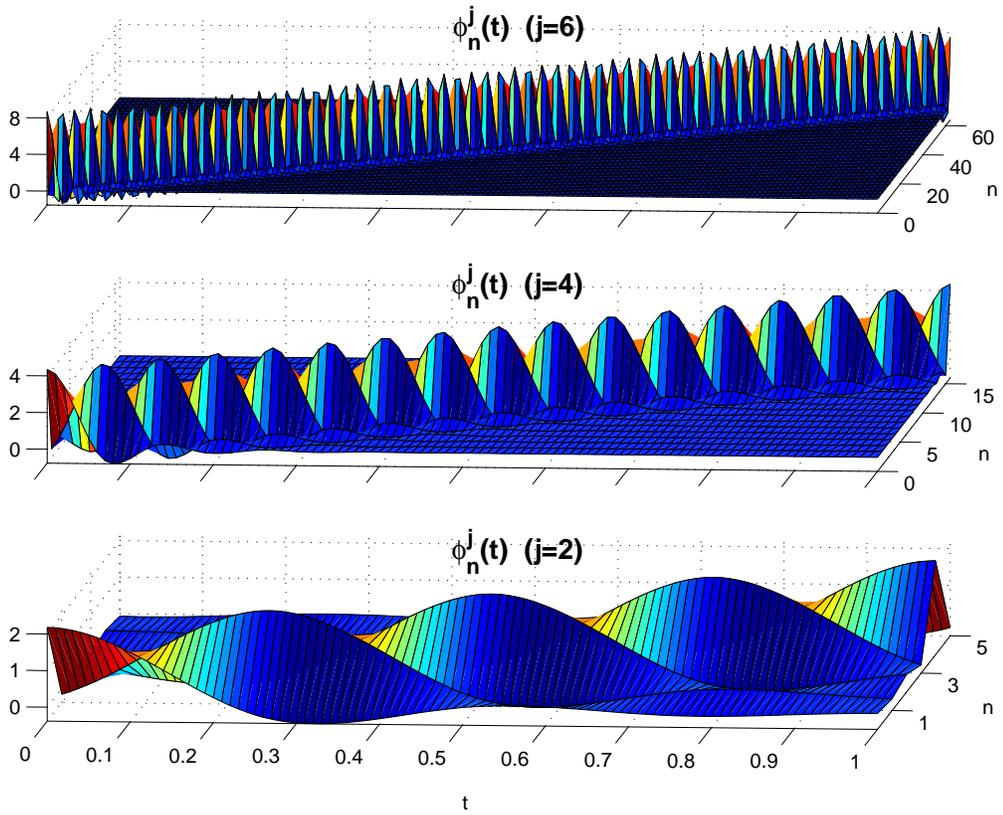}
   \caption
         {
	  $\phi_n^{j}$ as function of $t$ (physical space variable; here time) 
	  and $n$ (sampling space variable) for a selection of scale level $j$ 
	  (corresponding to scale $2^{-j}$ on a [0,1] domain).
	  The symmetric extension scheme is used in constructing 
	$\phi_n^j$ via the $\phi$ in Fig.~\ref{fig:scl_func}.
         \protect{\label{fig:scl_base}}}
   \end{center}
   \end{figure*}

Without loss of generality,
let the interval over which the signals to be diagnosed span be $[0,1]$;
if not, it may always be made so after a transformation. 
Consider a Hilbert space $V_{\ext,j} \subset L_2[0,1]$\footnote
   {Loosely speaking, it is a space of square integrable functions on
    [0,1].}
generated by the basis $\{\phi_n^j(t)\}_{n=0,1,...,2^j\ext-1}$, 
where 
	\begin{eqnarray}	\label{eq:scl_basis}
	\phi_n^{j}(t) = \sum_{q=-\infty}^{+\infty} 2^{j/2}
	\phi[2^j(t+\ext q)-n + 1/2], \qquad 
	   n=0,1,...,2^{j\ext-1}.
	\end{eqnarray}
Here $\phi(t)$ is a scaling function constructed in LA07
such that $\{\phi(t - n + 1/2)\}_{n}$ is orthonormal
(Fig.~\ref{fig:scl_func}). 
\trackchanges{
From $\phi(t)$ one can also construct an orthonormal wavelet basis.}
The parameter $\ext=1$ or $\ext=2$, 
corresponding respectively to the periodic and symmetric extension 
schemes. Shown in Fig.~\ref{fig:scl_base} is the basis for $\ext=2$
and a selection of $j$, namely, the ``scale level'' ($2^{-j}$ is the scale).
For notational simplicity, throughout this study 
the dependence of $\phi_n^j$ on $\ext$ is suppressed (but retained in 
other notations).

%
%	Figure fig:scl_func here
%	Figure fig:scl_base here
%
%

It has been justified in LA07 that there always exists a $j_2$ 
such that all the atmospheric/oceanic signals of concern lie 
in $V_{\ext,j_2}$. Furthermore, it has been shown in there that
	\begin{eqnarray*}
	V_{\ext,j_0} \subset V_{\ext,j_1} \subset V_{\ext,j_2},
	\qquad {\rm for}\ j_0 < j_1 <j_2.
	\end{eqnarray*}
A decomposition thus can be made such that
	\begin{eqnarray}
	V_{\ext,j_2} = V_{\ext,j_1} \oplus W_{\ext,j_1-j_2} 
	             = V_{\ext,j_0} \oplus W_{\ext,j_0-j_1}
				\oplus W_{\ext,j_1-j_2} 
	\end{eqnarray}
where $W_{\ext,j_1-j_2}$ is the orthogonal complement of $V_{\ext,j_1}$
in $V_{\ext,j_2}$, and $W_{\ext,j_0-j_1}$ that of
$V_{\ext,j_0}$ in $V_{\ext,j_1}$.
It has been shown by LA07 that $V_{\ext,j_0}$ contains functions of scales
larger than $2^{-j_0}$ only, while lying in $W_{\ext,j_0-j_1}$ and 
$W_{\ext,j_1-j_2}$ are the functions with scale ranges between 
$2^{-j_0}$ to $2^{-j_1}$ and $2^{-j_1}$ to $2^{-j_2}$, respectively.
We call the so-formed subspaces of $V_{\ext,j_2}$ as scale windows.
For easy reference, from larger scales (lower scale levels) to smaller scales
(higher scale levels), they will be referred to as scale windows 0, 1, and
2, respectively.
Depending on the problem of concern,
they may also be assigned names in association to physical processes.
For example, one may refer to them as large-scale, mid-scale, and
small-scale windows, or, in the context of, say, MJO studies, 
mean window, intraseasonal window or MJO window, and 
synoptic window, or, in the context of oceanography, large-scale window, 
meso-scale window and sub-mesoscale window.
More scale windows can be likewise defined,
but in this study, usually three are enough
(in fact in many cases only two are needed).

Consider a function $u(t) \in V_{\ext,j_2}$. With (\ref{eq:scl_basis}),
a transform
	\begin{eqnarray}
	\sclt u_n^j = \int_0^\ext u(t) \phi_n^j(t)\ dt,
	\end{eqnarray}
can be defined for a scale level $j$.
Given window bounds $j_0<j_1<j_2$, $u$ then can be reconstructed on the 
three scale windows as constructed above:
	\begin{eqnarray}
	\lgr u(t) 
	     &=& \sum_{n=0}^{2^{j_0}\ext-1} 
		\sclt u_n^{j_0} \phi_n^{j_0} (t), \label{eq:syn1} \\
	\mesor u(t) 
	     &=& \sum_{n=0}^{2^{j_1}\ext-1} \sclt u_n^{j_1} \phi_n^{j_1}(t) 
			- \lgr u(t), 	\label{eq:syn2}		\\
	\subr u(t) 
	     &=& u(t) - \lgr u(t) - \mesor u(t), 	\label{eq:syn3}
	\end{eqnarray}
with the notations $\sim$0, $\sim$1, and $\sim$2
signifying respectively the corresponding three scale windows.
Since $V_{\ext,j_0}$, $W_{\ext,j_0-j_1}$, $W_{\ext,j_1-j_2}$ are all
subspaces of $V_{\ext,j_2}$, the functions $\lgr u$, $\mesor u$, $\subr u$ 
can be transformed with respect to $\{\phi_n^{j_2}(t)\}_n$,
the basis of $V_{\ext,j_2}$,
	\begin{eqnarray}		\label{eq:mwt}
	\swt u n \win = \int_0^\ext \swr u \win (t)\ 
		        \phi_n^{j_2}(t) \ dt,
	\end{eqnarray}
for windows $\win=0,1,2$, and $n=0,1,...,2^{j_2\ext}-1$.
Note here the transform coefficients $\swt u n \win$ contains only the
processes belonging to scale window $\win$.
It has, though discretely, the finest resolution permissible 
in the sampling space on $[0,1]$.
We call (\ref{eq:mwt}) a {\it multiscale window transform}, or MWT for short. 
With this, (\ref{eq:syn1}), (\ref{eq:syn2}), and (\ref{eq:syn3})
can be written in a unified way:
		\begin{eqnarray}		\label{eq:mwr}
		\swr u \win (t) = \sum_{n=0}^{2^{j_2}\ext-1} \swt u n \win
				\phi_n^{j_2}(t), \qquad \win=0,1,2.
		\end{eqnarray}
Eqs.~(\ref{eq:mwt}) and (\ref{eq:mwr}) form the 
transform-reconstruction pair for MWT.

\subsection{Multiscale energy}		\label{sect:ms_energy}

MWT has a Parseval relation-like property; in the periodical extension case 
($\ext=1$),
		\begin{eqnarray}	\label{eq:marginalization}
		\sum_n \swt u n \win \swt v n \win  
		= \overline{\swr u \win (t)\ \swr v \win (t)},
		\end{eqnarray}
for  $u,v\in V_{1,j_2}$, and because of the mutual orthogonality between
the scale windows,
		\begin{eqnarray}	\label{eq:marg2}
		\sum_\win \sum_n \swt u n \win \swt v n \win
		= \overline{u(t) v(t)},
		\end{eqnarray}
where the overline indicates averaging over time, and
$\sum_n$ is a summation over the sampling set $\{0,1,2,...,2^{j_2}-1\}$ 
(see LA07 for a proof). In the case of other extensions, 
$\sum_n$ is replaced by ``marginalization'', a naming
convention after 
	 \cite{Huang},
	% Huang et al. (1999), 
which also bears the physical meaning of summation over $n$.
Eq.~(\ref{eq:marg2}) states that, 
a product of two MWT coefficients followed by 
a marginalization is equal to the product of their 
corresponding reconstructions averaged over the duration. 
This property is usually referred to as {property of marginalization}.

The property of marginalization is important in that it 
allows for an efficient representation of multiscale energy in terms 
of the MWT transform coefficients.
In (\ref{eq:marg2}), let $u=v$, 
the right hand side is then the energy of $u$ (up to some constant factor)
averaged over $[0,1]$. It is equal to a summation of $3N= 3 \times 2^{j_2}$ 
(if 3 scale windows are considered)
individual objects $\parenth{\swtnw u}^2$ centered at time 
$t_n = 2^{-j_2}n+ 1/2$, with a characteristic influence interval 
$\Delta t = t_{n+1} - t_n = 2^{-j_2}$.
The multiscale energy at time $t_n$ then should be the mean 
over the interval:
	$\frac {\parenth{\swtnw u}^2} {\Delta t}
		 = 2^{j_2} {\parenth{\swtnw u}^2}.$
Notice the constant multiplier $2^{j_2}$; it is needed for the obtained
multiscale energy to make sense in physics.
But for notational succinctness, it will be omitted in the following
derivations.

Therefore, the energy of $u$ on scale window $\win$ at step $n$ is
	\begin{eqnarray}
	E_n^\win \propto \parenth{\swtnw u }^2.
	\end{eqnarray}
Note the $\win$-window filtered signal is $\swr u \win$; by the common
practice one would take $\parenth{\swr u \win}^2$ as the energy on scale 
window $\win$. From above one sees that this is conceptually incorrect.

\section{Multiscale flux and canonical transfer} \label{sect:separation}

\subsection{Multiscale flux}	\label{sect:flux}
For a scalar field $T$, its ``energy'' (quadratic property) 
on window $\win$ at step $n$
is $\frac 12 (\swtnw T)^2$ (up to some factor). 
In the MWT framework, energy can be decomposed as a sum of a bunch of 
atom-like elements:
	\begin{eqnarray}
	\frac12 T^2 = \sum_{n_1,\win_1} \sum_{n_2,\win_2} \frac12
		\bracket{\swt T {n_1} {\win_1} \phi_{n_1}^{j_2}(t)}
		\bracket{\swt T {n_2} {\win_2} \phi_{n_2}^{j_2}(t)}.
	\end{eqnarray}
Look at the flux of the ``atom'' by a flow $\ve v(t)$ over
$t\in[0,1]$ at step $n$ within window $\win$. It is
	\begin{eqnarray}
	\int_0^1 \ve v(t) \cdot \frac12 
		\bracket{\swt T {n_1} {\win_1} \phi_{n_1}^{j_2}(t)}
		\bracket{\swt T {n_2} {\win_2} \phi_{n_2}^{j_2}(t)}
		\cdot
		\delta(n-n_2) \cdot \delta(\win - \win_2)\ dt.
	\end{eqnarray}
% Note the integration over the time series is not just needed for the MWT
% purpose; it is also necessary from the energetic point of view. Generally
% speaking, an energetic process is not a local concept. It is, however,
% possible to make it spatially local by performing a temporal integration,
% provided that the process is ergodic. In this sense, a localized energetic
% formalism within the framework of MWT is indeed the natural choice.
%
In the above delta functions, the arguments may equally be chosen as $n_1$
and $\win_1$. The flux of $\frac12 T^2$ by the flow $\ve v$ on $\win$ at step
$n$ is then the sum of the atomic expressions over all the possible $n_1$,
$n_2$, $\win_1$, and $\win_2$, i.e.,
	\begin{eqnarray}
	\ve Q_n^\win 
	&=& \sum_{n_1,\win_1} \sum_{n_2,\win_2} 
		\int_0^1 \frac12 \ve v \cdot
		\bracket{\swt T {n_1} {\win_1} \phi_{n_1}^{j_2}(t)}
		\bracket{\swt T {n_2} {\win_2} \phi_{n_2}^{j_2}(t)}
		\cdot
		\delta(n-n_2) \delta(\win - \win_2) dt	\cr
	&=&
	    \frac12 \int_0^1 \ve v(t) T(t) \cdot \swtnw T 
		\phi_n^{j_2}(t) dt.
	\end{eqnarray}
But the function $\swtnw T \phi_n^{j_2}(t)$ lies in window $\win$, and
all windows are orthogonal, so this is something like a projection of
$\ve v T$ onto window $\win$:
	\begin{eqnarray}	\label{eq:ms_Q}
	\ve Q_n^\win 
	&=& \frac12 \int_0^1 \swtnw{(\ve vT)}  \cdot
			\swtnw T \phi_n^{j_2}(t) dt	\cr
	&=& \frac12 \swtnw T \swtnw {(\ve v T)}.
	\end{eqnarray}

The above can be used for the derivation of multiscale potential
energetics. For kinetic energy $K = \frac12 \ve v \cdot \ve v$, essentially
one can derive in the same way. To avoid confusion, we consider the
energy-like quantity of an arbitrary vector $\ve G$, 
	\begin{eqnarray}
	K = \frac 12 \ve G\cdot \ve G
	=\sum_{n_1,\win_1} \sum_{n_2,\win_2} \frac12
		\bracket{\swt {\ve G} {n_1} {\win_1} \varphi_{n_1}^{j_2}(t)}
		\cdot
		\bracket{\swt {\ve G} {n_2} {\win_2} \varphi_{n_2}^{j_2}(t)}.
	\end{eqnarray}
So the flux of the ``atom'' over $t\in[0,1]$ at step $n$ on window $\win$
is
	\begin{eqnarray}
	\int_0^1 \ve v(t) \ \frac12 
		\bracket{\swt {\ve G} {n_1} {\win_1} \varphi_{n_1}^{j_2}(t)}
		\cdot
		\bracket{\swt {\ve G} {n_2} {\win_2} \varphi_{n_2}^{j_2}(t)}
		\
		\delta(n-n_2) \delta(\win-\win_2) dt,
	\end{eqnarray}
and the flux of $K$ by $\ve v$ on $\win$ at $n$ is
	\begin{eqnarray}
	\ve Q_n^\win 
	&=&
	    \sum_{n_1,\win_1} \sum_{n_2,\win_2} \frac12 \ve v(t)
		\bracket{\swt {\ve G} {n_1} {\win_1} \varphi_{n_1}^{j_2}(t)}
		\cdot
		\bracket{\swt {\ve G} {n_2} {\win_2} \varphi_{n_2}^{j_2}(t)}
		\
		\delta(n-n_2) \delta(\win-\win_2) dt	\cr
	&=& \frac12 \int_0^1 \bracket{\ve v(t) \ve G(t)}
			\cdot \swtnw {\ve G} \varphi_n^{j_2}(t) dt,
	\end{eqnarray}
where the dyadic $\ve v \ve G$ takes right dot product with $\swtnw {\ve
G}$. Again, $\swtnw {\ve G} \varphi_n^{j_2}(t)$ lies in window $\win$. Due
to the orthogonality among windows,
	\begin{eqnarray}
	\ve Q_n^\win 
	&=& \frac12 \int_0^1 \bracket{\ve v(t) \ve G(t)}^{\sim\win}
		\cdot \swtnw {\ve G} \varphi_n^{j_2}(t) dt	\cr
	&=& \frac12 \swtnw {(\ve v \ve G)} \cdot \swtnw {\ve G}	\cr
	&=& \frac12 \bracket{
		  \swtnw{(\ve v G_1)} \swtnw{(G_1)}
		+ \swtnw{(\ve v G_2)} \swtnw{(G_2)}
			    },
	\end{eqnarray}
which is like the superposition of the fluxes of two scalar fields, namely,
$G_1$ and $G_2$.

\subsection{Canonical transfer}		\label{sect:canonical}

Consider a scalar property $T$ in an incompressible flow field $\ve v$.
The equation governing the evolution of $T$ is
	\begin{eqnarray*}
	\Dt T + \grad\cdot (\ve v T) = {\rm other\ terms}.
	\end{eqnarray*}
As only the nonlinear term namely the advection will lead to interscale
transfer, all other terms (e.g., diffusion, source/sink) 
are unexpressed and put to the right hand side. 
To find its evolution on window $\win$, take MWT on both sides. The first
term is $\swtnw {\parenth{\Dt T}}$. It has been shown by LR05 to be
approximately equal to $\frac {\delta {\swtnw T}} {\delta n}$,
	% (the difference is negligible), 
where 
$\frac \delta  {\delta n}$ is the difference operator 
with respect to $n$. Since $t$ of the physical space 
is now carried over to $n$ of the sampling space, 
the difference operator is essentially the time rate of change when
applying to a discrete time series. 
We therefore would write it as $\Dt {\swtnw T}$ to avoid introducing
extra notations, which are already too many.
But the careful reader should bear in mind that here it means the
difference in the sampling space rather than the differential in the
physical space. (Since the signals are sampled at each time step, in
real applications they are precisely the same.) 
The MWTed equation is, therefore,
	\begin{eqnarray*}
	\Dt {\swtnw T} + \grad \cdot \swtnw {(\ve v T)} = ...
	\end{eqnarray*}
Multiplication of $\swtnw T$ gives
	\begin{eqnarray}	\label{eq:E}
	\Dt {E_n^\win} = -\swtnw T \nabla \cdot \swtnw{(\ve v T)} + ...
	\end{eqnarray}
where $E_n^\win = \frac12 \parenth{\swtnw T}^2$ 
is the energy on window $\win$ at step $n$.

One continuing effort in multiscale energetics study is to separate
	$-\swtnw T \nabla \cdot \swtnw{(\ve v T)}$
into a transport process term ($\grad\cdot\ve Q_n^\win$) 
and a transfer process term ($\Gamma_n^\win$). 
Symbolically this is
	\begin{eqnarray*}
	-\grad\cdot\ve Q_n^\win + \transfer_n^\win.
	\end{eqnarray*}
An intuitively and empirically based common practice is to
collect divergence terms to form the transport term 
	 (e.g., \cite{Harrison}; \cite{Pope}).
	% (e.g., Harrison and Robinson, 1978; Pope, 2003). 
However, as long pointed by people such as 
	 \cite {Holopainen}, \cite{Plumb83},
	% Holopainen (1978), 
	% Plumb (1983),
among others, there exist other forms that may result in different
separations. 

In this study, the separation is natural. 
The multiscale flux $\ve Q_n^\win$, hence the multiscale transport, 
has been rigorously obtained in the preceding subsection 
[i.e., Eq.(\ref{eq:ms_Q})]!
The transfer $\transfer$ is obtained by 
subtracting $-\nabla\cdot\ve Q_n^\win$ from the right hand side of 
(\ref{eq:E}): 
	\begin{eqnarray}		\label{eq:transfer}
        \transfer_n^\win 
	&=& - \swtnw T \nabla \cdot \swtnw {(\ve v T)}
            + \nabla \cdot \bracket{\frac 1 2 \swtnw T \swtnw{(\ve v T)}} \cr
	&=& \frac 1 2 \bracket{
		\swtnw{(\ve v T)} \cdot \nabla \swtnw T -
		\swtnw T \nabla \cdot \swtnw{(\ve v T)}
				}.
	\end{eqnarray}
Notice that the resulting transfer bears a form similar to the Lie bracket and,
particularly, the Poisson bracket in Hamiltonian mechanics. 
\trackchanges{
To see this, recall that a Poisson bracket $\{\cdot,\cdot\}$ is defined,
for differential operators ($\Dx\ $,~$\Dy\ $) and
functions $F$ and $G$, such that
	\begin{eqnarray*}
	\{F,G\} = \Dy F \Dx G - \Dx F \Dy G.
	\end{eqnarray*}
If $\{F,G\} = 0$, $F$ and $G$ are said to be involution or to Poisson commute.
Consider the 1D version of $\transfer_n^\win$, i.e.,
	\begin{eqnarray*}
	\frac 1 2 \bracket{
		\swtnw{(u T)} \Dx {\swtnw T} -
		\swtnw T \Dx {\swtnw{(u T)}}
				}.
	\end{eqnarray*}
If we pick two differential operators $(\Dx\ , \mathbb I)$, where 
$\mathbb I$ is the identity, then the above canonical transfer is
simply $\frac12 \{\swtnw{(uT)}, \swtnw T\}$.
} Because of this, we will refer it to as {\it canonical transfer} in the
future, in order to distinguish it from other transfers already existing 
in the literature.

Canonical transfers possess a very important property,
as stated in the following theorem.
	\begin{thm}
	A canonical transfer vanishes upon summation over all
	the scale windows and marginalization over the sampling space,
	i.e.,
	\begin{eqnarray}	\label{eq:canonical}
	\sum_n \sum_\win \transfer_n^\win = 0. 
	\end{eqnarray}
	\end{thm}
%
%    \begin{center}
%      \framebox[0.3\textwidth]{
%      \begin{minipage}	[c]{1\textwidth}
%	\begin{eqnarray}	\label{eq:canonical}
%	\qquad \sum_n \sum_\win \transfer_n^\win = 0. \\
%					\nonumber
%	\end{eqnarray}
%      \end{minipage}
%      }
%    \end{center}
{\bf Remark}: 
This theorem states that a canonical transfer process only re-distributes 
energy among scale windows, without generating or destroying energy as a whole.
This is precisely that one would expect for an energy transfer process!
This property, though natural, generally does not hold for the existing 
empirical formalisms.

\pf
By the property of marginalization (\ref{eq:marginalization}), 
Eq.~(\ref{eq:transfer}) gives
	\begin{eqnarray*}
	\sum_n \transfer_n^\win =
	\frac 1 2 \int_0^1 \bracket{
		  \swr {(\ve v T)} \win \cdot \nabla \swr T \win - 
		  \swr T \win \nabla\cdot \swr {(\ve v T)} \win}\
		dt.
	\end{eqnarray*}
Because of the orthogonality between different scale windows, this followed 
by a summation over $\win$ results in
	\begin{eqnarray*}
	\frac 1 2 \int_0^1 \bracket{
		(\ve  v T) \cdot \nabla T -
		T \nabla\cdot (\ve v T)
				} \ dt = 0.
	\end{eqnarray*}
In the above derivation, the incompressibility assumption of the flow 
has been used.

The canonical transfer (\ref{eq:transfer}) may be further simplified in
expression when $\swtnw T$ is nonzero:
	\begin{eqnarray}
	\transfer_n^\win &=& 
	      - E_n^\win\ \nabla \cdot \parenth
		{\frac {\swtnw {(\ve v T)}} {\swtnw T}}, \
		{\rm if\ }\swtnw T \ne 0,	\label{eq:svelo}	
	\end{eqnarray}
where $E_n^\win = \frac 1 2 \parenth{\swtnw T}^2$ is the energy on window $\win$
at step $n$, and is hence always positive.
Note that (\ref{eq:svelo}) defines a field variable which has the
dimension of velocity in physical space: 
	\begin{eqnarray}
	\ve v^\win_T = \frac {\swtnw {(T \ve v)}} {\swtnw T}.
	\end{eqnarray}
It may be loosely understood as a weighted average of $\ve v$, 
with the weights derived from the MWT of the scalar field $T$. 
For convenience, we will refer to $\ve v^\win_T$ as {\it $T$-coupled velocity}.
The growth rate of energy on window $\win$ 
is now totally determined by 
	$-\nabla\cdot \ve v^\win_T$, the convergence of $\ve v^\win_T$,
and
	\begin{eqnarray}	\label{eq:svelo2}
	\transfer_n^\win = - E_n^\win \nabla \cdot \ve v^\win_T.
	\end{eqnarray}
Note $\transfer_n^\win$ makes sense even when $\swtnw T=0$ and hence 
$\ve v^\win_T$ does not exist. In this case, (\ref{eq:svelo2}) should be 
understood as (\ref{eq:transfer}).
% We may keep using (\ref{eq:svelo}) and (\ref{eq:svelo2}) for notational 
% simplicity and physical clarity.

\trackchanges{
The canonical transfer has been validated in many applications.
Particularly, it verifies the barotropic instability structure of the Kuo
jet stream model which fails the classical empirical formalism.
To facilitate the comparison, Liang and Robinson (2007)\cite{LR07} established that,
when $j_0=0$ and a periodical
extension is used, the canonical transform (\ref{eq:transfer}) 
is reduced to
	\begin{eqnarray*}
	\frac 12 [\bar T \grad\cdot \overline{\ve v'T'}
		 - \overline{\ve v'T'} \cdot \grad\bar T]
	\end{eqnarray*}
(the overbar indicates a time mean over the whole duration),
which is also in a Lie bracket form. 
This is quite different from the traditional transfer 
$-\overline{\ve v'T'}\cdot \grad\bar T$, which, when $T$ is a component of velocity, 
is usually understood as the energy extracted by the Reynolds stress 
against the basic profile $\bar T$. As demonstrated in \cite{LR07}, this
``Reynolds extraction'' does not verify the analytical solution of 
the Kuo instability model, while our canonical transfer does.

}

\section{Multiscale atmospheric energetics}	\label{sect:atmos}
We now apply the above theory to derive the multiscale atmospheric energetics.
For notational brevity, from now on the {\it dependence on $n$ 
will be suppressed} in the MWT terms, unless otherwise indicated.

\subsection{Primitive equations}

Consider an ideal gas and assume hydrostaticity to hold. We adopt an
isobaric coordinate system, which is advantageous over others in that
air may be viewed as incompressible, and, besides, as we will see,
the resulting energy equations are free of density. The governing equations
are (see, for example, \cite{Salby}):	% Salby, 1996
	\begin{eqnarray}
	&& \Dt{\vh} + \vh \cdot \gradh\vh + \w \Dp\vh 
			+ f\ve k \times \ve \vh
	    = -\gradh\gp^* + \ve F_{m,p} + \ve F_{m,h},   
						\label{eq:atmos_gov1a} \\
	&& \Dp\gp^* = -\alpha^*,		\label{eq:tmos_gov2a} \\
	&& \gradh\cdot\vh + \Dp\w = 0,		\label{eq:atmos_gov3a} \\
	&& \Dt {T^*} + \vh \cdot \gradh T^* + \w\Dp {T^*}
		- \frac{\alpha^*\w} {c_p}
		= \frac {\dot q_{net}} {c_p},	\label{eq:atmos_gov4a} \\
	&& p\alpha^* = RT^*,			\label{eq:atmos_gov5a}
	\end{eqnarray}
where $\dot q_{net}$ stands for the heating rate from all diabatic 
sources, $\w = \dt p$, and the starred variables mean the whole fields
\trackchanges{(do not include velocity)}, 
with the corresponding non-starred ones reserved for their anomalies.
The subscript $h$ indicates the component on the $p$ plane; 
for example, $\ve v = (\vh, \w)$, $\grad = (\gradh, \Dp\ )$, and 
so forth. The other symbols are conventional (cf.~Appendix~A).

%	\begin{eqnarray}
%	&& \Dt{\vh} + \frac u {a\cos\lat} \Dlon {\vh}
%		 + \frac va \Dlat {\vh}
%		 + \w \Dp {\vh}
%		 + \parenth{f + u \frac {\tan\lat} a} \ve k \times \vh \cr
%	&&\qquad\qquad
%	    = -\gradh\gp + \ve F_{m,p} + \ve F_{m,h},   \label{eq:gov1} \\
%	&& \Dp\gp = -\alpha,				\label{eq:gov2} \\
%	&& \gradh\cdot\vh + \Dp\w = 0,			\label{eq:gov3} \\
%	&& \Dt T + \frac u {a\cos\lat}\Dlon T
%		 + \frac v a \Dlat T  +  \w\Dp T
%				    + \w \bar\alpha \frac {L-L_d}g
%	                            + \w \alpha \frac {L-L_d}g
%		= \frac{\dot q_{net}}{c_p}.		\label{eq:gov4}
%	\end{eqnarray}

% We start off the derivation with the temperature equation 
% (\ref{eq:atmos_gov4a}).

Let $\bar T$ denote the
temperature averaged over the $p$-plane and time, and $T$ the departure of
$T^*$ from $\bar T$. Then
	\begin{eqnarray}
	T^* = \bar T(p) + T(\lon,\lat,p;t).
	\end{eqnarray}
The ideal gas law (\ref{eq:atmos_gov5a}), or 
	$\alpha^* = \frac R p T^*,$
implies a linear relation between $T$ and $\alpha$, and hence, equally, we
have
	\begin{eqnarray}
	\alpha^* = \bar \alpha(p) + \alpha(\lon,\lat,p;t).
	\end{eqnarray}
By hydrostaticity
	\begin{eqnarray}
	\gp^* = \bar\gp(p_s) - \int_{p_s}^p \alpha^* dp 
	      = \bar\gp(p_s) - \int_{p_s}^p \bar\alpha dp 
			     - \int_{p_s}^p \alpha dp
	      \equiv \bar\gp(p) + \gp.
	\end{eqnarray}
%
%	\begin{figure}	[h]
%	\begin{center}
%	\includegraphics[angle=0, width=0.25\textwidth]
%	{figures/grid_p.eps}
%	\caption{
%	Grid in $p$ coordinate. In the figure, 
%	$\delp(k)$ is actually $\delp(k-\frac12)$.
%	\protect{\label{fig:grid_p}}}
%	\end{center}
%	\end{figure}

The heat equation (\ref{eq:atmos_gov4a}) may then be re-written in terms of
$T$:
	\begin{eqnarray*}
	\Dt T + \vh \cdot\gradh T + \w\Dp T + \w\Dp {\bar T} 
		- \w \frac {\alpha^*} {c_p} = \frac{\dot q_{net}}{c_p}.
	\end{eqnarray*}
But
	% \begin{linenomath*}
	\begin{eqnarray*}
	\frac 1 {\alpha^*} \Dp{\bar T}
	= -\frac 1 g \Dz p \cdot \Dp {\bar T} = -\frac1g \Dz {\bar T}
	= \frac 1g L,
	\end{eqnarray*}
	% \end{linenomath*}
where $L = - \Dz {\bar T}$ is the Lapse rate.  
	% ($\approx 6.5\times 10^{-3} K/m$ in troposphere?). 
Also let
	\begin{eqnarray*} 
	L_d \equiv \frac g {c_p} \approx 9.8\times 10^{-3} K/m
	\end{eqnarray*}
(lapse rate for dry air). The above equation hence becomes
	\begin{eqnarray}	\label{eq:temp2}
	\Dt T + \vh \cdot \gradh T + \w\Dp T + \w \alpha^* \frac {L-L_d}g
		= \frac{\dot q_{net}}{c_p}.
	\end{eqnarray}
Note that
	\begin{eqnarray*}
  \alpha^* \frac {L-L_d} g = \alpha^* \parenth{-\Dz{\bar T} - \frac 1{c_p}}
	= \frac {R \bar T} {c_p p} - \Dp {\bar T}
	= -\frac{\bar T}\theta \Dp \theta \equiv S_p  
	\end{eqnarray*}
is the stability parameter ($\theta$ is the potential temperature).

From above we also have
	\begin{eqnarray}
	\gradh \gp^* = \gradh \gp,
	\end{eqnarray}
and by the hydrostatic assumption,
	\begin{eqnarray}
	\Dp\gp = \Dp {\gp^*} - \Dp {\bar\gp} 
		= -\alpha^* + \bar\alpha = -\alpha.
	\end{eqnarray}
Hence the primitive equations are, in term of $T$, $\gp$, etc.,
	\begin{eqnarray}
	&& \Dt{\vh} + \vh\cdot\gradh\vh + \w\Dp\vh + f\ve k\times \vh
	    = -\gradh\gp + \ve F_{m,p} + \ve F_{m,h}, \label{eq:atmos_gov1}\\
	&& \Dp\gp = -\alpha,			\label{eq:atmos_gov2} \\
	&& \gradh\cdot\vh + \Dp\w = 0,		\label{eq:atmos_gov3} \\
	&& \Dt T + \vh\cdot\gradh T + \w\Dp T
		    + \w \bar\alpha \frac {L-L_d}g
                    + \w \alpha \frac {L-L_d}g
		= \frac{\dot q_{net}}{c_p}, 	\label{eq:atmos_gov4} \\
	&& \alpha = \frac R p T			\label{eq:atmos_gov5}
	\end{eqnarray}
%
%	\begin{eqnarray}
%	&& \Dt{\vh} + \frac u {a\cos\lat} \Dlon {\vh}
%		 + \frac va \Dlat {\vh}
%		 + \w \Dp {\vh}
%		 + \parenth{f + u \frac {\tan\lat} a} \ve k \times \vh \cr
%	&&\qquad\qquad
%	    = -\gradh\gp + \ve F_{m,p} + \ve F_{m,h},   \label{eq:gov1} \\
%	&& \Dp\gp = -\alpha,				\label{eq:gov2} \\
%	&& \gradh\cdot\vh + \Dp\w = 0,			\label{eq:gov3} \\
%	&& \Dt T + \frac u {a\cos\lat}\Dlon T
%		 + \frac v a \Dlat T  +  \w\Dp T
%				    + \w \bar\alpha \frac {L-L_d}g
%	                            + \w \alpha \frac {L-L_d}g
%		= \frac{\dot q_{net}}{c_p}.		\label{eq:gov4}
%	\end{eqnarray}
In the heat equation	
$\w \alpha \frac {L-L_d}g$ makes a correction term and 
is by comparison small (since $\alpha \ll \bar\alpha$).
%
% In the above equations, the following notations may be occassionally 
% used to simplify the expressions
%	\begin{eqnarray}
%	&& \dt\ = \Dt\ + \frac u {a\cos\lat}\Dlon\ 
%		 + \frac v a \Dlat\   +  \w\Dp\ ,	\\
%	&& \D_v = \frac u {a\cos\lat}\Dlon\ 
%		 + \frac v a \Dlat\   +  \w\Dp\ .
%	\end{eqnarray}

\subsection{Multiscale kinetic energy equations}

The start step is to find $\ve Q_{K,n}^\win$, 
the flux on scale window $\win$ at step $n$. 
This has been fulfilled in the preceding section,
which we rewrite here for reference,
	\begin{eqnarray}
	\ve Q_K^\win = \frac12 \widemwtn {\ve v \vh} \cdot \vhmwtn.
	\end{eqnarray}
Componentwise this is
	\begin{eqnarray}
	Q_{K,\lon}^\win &=& \frac12 \bracket{\widemwtn {uu} \mwtn u 
			+ \widemwtn{uv} \mwtn v},	\label{eq:Q1}\\
	Q_{K,\lat}^\win &=& \frac12 \bracket{\widemwtn {vu} \mwtn u 
			+ \widemwtn{vv} \mwtn v},	\label{eq:Q2}\\
	Q_{K,p}^\win &=& \frac12 \bracket{\widemwtn {\w u} \mwtn u 
			+ \widemwtn{\w v} \mwtn v}.	\label{eq:Q3}
	\end{eqnarray}
From the horizontal momentum equations, the canonical transfer is
	\begin{eqnarray*}
	\Gamma_K^\win 
	&=& - \widemwtn {\ve v \cdot \grad \vh} \cdot \vhmwtn
			+ \grad \cdot \ve Q_{K}^\win.
%	&=& - \widemwtn {\D_v \vh} \cdot \vhmwtn 
%	    - \widemwtn {u \frac {\tan\lat} a \ve k \times \vh} \cdot\vhmwtn
%	    + \grad \cdot \ve Q_n^\win. 
	\end{eqnarray*}
It is better expressed, with the aid of the incompressibility
equation (\ref{eq:atmos_gov3}), as
	\begin{eqnarray}
	\Gamma_K^\win 
	&=& -\bracket{\grad\cdot\widemwtn{\ve v \vh}} \cdot \vhmwtn
	    + \grad \cdot \ve Q_{K}^\win,          \label{eq:Gamma1}  \\
	&=&
	    -\bracket{\grad\cdot\widemwtn{\ve v \vh}} \cdot \vhmwtn
	    + \frac12 \grad \cdot \bracket{\widemwtn{\ve v \vh} 
		\cdot \vhmwtn}		\cr
	&=& 
	    \frac12 \left\{
		\widemwtn {\ve v\vh} : \grad\vhmwtn
		- \bracket{\grad \cdot \widemwtn {\ve v\vh}} \cdot \vhmwtn
		    \right\},
	\end{eqnarray}
where the colon operator $:$ is defined such that, for two dyadic products
$\ve A \ve B$ and $\ve C \ve D$,
	\begin{eqnarray*}
	(\ve A \ve B) : (\ve C \ve D) = 
	(\ve A \cdot \ve C) (\ve B \cdot \ve D).
	\end{eqnarray*}

In fact, the above can be expanded in terms of the components of 
$\vh = (u,v)$, i.e.,
	\begin{eqnarray}
	\Gamma_K^\win = 
	   \frac12 \cbrace{\widemwtn{\ve v u} \grad\mwtn u
			- [\grad\cdot\widemwtn{\ve v u}] \mwtn u}
	   +
	   \frac12 \cbrace{\widemwtn{\ve v v} \grad\mwtn v
			- [\grad\cdot\widemwtn{\ve v v}] \mwtn v}.
	\end{eqnarray}
Notice that this is just the sum of two canonical transfers, and is hence 
canonical.

The equation governing the evolution of $K^\win = \frac12 \vhmwtn \cdot
\vhmwtn$ is, therefore
[after dot multiplying $\vhmwtn$ with the MWT of (\ref{eq:atmos_gov1})], 
	\begin{eqnarray}
	\Dt {K^\win} + \grad\cdot\ve Q_{K}^\win 
	&=& \Gamma_K^\win - \grad\cdot 
			  \parenth{\mwtn{\ve v} \mwtn\gp }
			- \mwtn\w \mwtn\alpha 
			+ F_{K,p}^\win + F_{K,h}^\win	\cr
	&=& \Gamma_K^\win - \grad\cdot \ve Q_{P}^\win
			  - b^\win
			+ F_{K,p}^\win + F_{K,h}^\win.
				\label{eq:ke_atmos}
	\end{eqnarray}	
Here $\ve Q_{P}^\win = \mwtn {\ve v} \mwtn\gp$,
and $b^\win = \mwtn\w \mwtn\alpha$ is the rate of buoyancy conversion.

It is necessary to derive the expressions in spherical coordinates.
\trackchanges{
If the vertical coordinate is $z$, the Lam\'e's coefficients are
	$h_\lon \approx a\cos\lat,
	 h_\lat \approx a,
	 h_z = 1,$
where $a$ is the radius of Earth, and $(\lon,\lat)$ are longitude and latitude,
thus the divergence of $\ve Q_K^\win = 
	(Q_{K,\lon}^\win, Q_{K,\lat}^\win, Q_{K,z}^\win)$ 
is
	\begin{eqnarray}	\label{eq:divQ}
	\grad\cdot\ve Q 
	&=& \frac 1 {h_\lon h_\lat h_z} \bracket{
		\Dlon {(h_\lat h_z Q_{K,\lon}^\win)} 
		+ \Dlat {(h_\lon h_z Q_{K,\lat}^\win)} 
		+ \Dz{(h_\lon h_\lat Q_{K,z}^\win)} 
						      } \cr
	&=& \frac 1 {a\cos\lat} \Dlon {Q_{K,\lon}^\win} +
	    \frac 1 {a\cos\lat} \Dlat {(Q_{K,\lat}^\win \cos\lat)} + 
	    \Dz {Q_{K,z}^\win}.
	\end{eqnarray}
If the vertical coordinate is $p$, $\grad\cdot\ve Q_K^\win$ can also
be approximately expressed as, 
	\begin{eqnarray}	\label{eq:Q_spher}
	\grad\cdot\ve Q_{K}^\win
	= \frac 1 {a\cos\lat} \Dlon {Q_{K,\lon}^\win} +
	  \frac 1 {a\cos\lat} \Dlat {(Q_{K,\lat}^\win \cos\lat)} +
	  \Dp {Q_{K,p}^\win}.
	\end{eqnarray}
The components of $\ve Q_{K}^\win$ are referred to
(\ref{eq:Q1})-(\ref{eq:Q3}).}
Note that this is just an approximate expression, as this is not strictly
an orthogonal frame. 
  % the metric tensor has off-diagonal entries, (2) the
  % Lam\'e coefficient $h_3 =  -\frac 1 {\rho g} = - \frac\alpha g$ 
  % is not a constant. 
However, since the shell of the atmosphere is thin
(shallow-water assumption), the $p$ direction may be viewed as unaffected,
just as in the geographic coordinate system. Likewise,
	\begin{eqnarray}
	\grad\cdot\ve Q_{P}^\win
	= \frac 1 {a\cos\lat} \Dlon {(\mwtn u \mwtn\gp)} +
	  \frac 1 {a\cos\lat} \Dlat {(\mwtn v \mwtn\gp \cos\lat)} +
	  \Dp {(\mwtn\w \mwtn\gp)}.
	\end{eqnarray}
The difficulty is with the transfer term. It would be easier to start from
(\ref{eq:Gamma1}). By the result of Appendix~C,
	\begin{eqnarray}
	&& \grad\cdot (\ve v \ve v) 
	=
	\cbrace{
	   \frac1{a\cos\lat}\bracket{
		 \Dlon{u^2} - uv \sin\lat + u\w \cos\lat
		+ \Dlat {(vu\cos\lat)}
				    }  + \Dp {\w u}
		          } \ve e_\lon		\cr
	&&\quad +
	\cbrace{
	   \frac1{a\cos\lat}\bracket{
		 \Dlon{uv} + u^2 \sin\lat + \Dlat {(v^2\cos\lat)}
			+ v\w \cos\lat
				    }  + \Dp {\w v}
		          } \ve e_\lat		\cr
	&&\quad +
	\cbrace{
	   \frac1{a\cos\lat}\bracket{
		 \Dlon{u\w} - u^2 \cos\lat + \Dlat {(v\w\cos\lat)}
			- v^2 \cos\lat
				    }  + \Dp {\w^2}
		          } \ve e_p.	\label{eq:gradvv2}
	\end{eqnarray}
In particular,
	\begin{eqnarray*}
	\grad\cdot(\ve v\vh)
	&=& \left\{ \frac 1 {a\cos\lat}  
	        \bracket{\Dlon{u^2} - uv \sin\lat + \Dlat {uv\cos\lat}} 
		+ \Dp{\w u}
	    \right\} \ve e_\lon			\\
	&& +   \left\{ \frac 1 {a\cos\lat}
		\bracket{\Dlon {uv} + u^2 \sin\lat + \Dlat {v^2\cos\lat}}
		+ \Dp {\w v}
	    \right\} \ve e_\lat 		\\
	&& -   \left\{  
	         \frac {u^2 + v^2} a
	    \right\} \ve e_p.
	\end{eqnarray*}
So
	\begin{eqnarray}
	\Gamma_K^\win 
	&=& -\bracket{\grad\cdot \widemwtn{\ve v\vh}} \cdot\vhmwtn
			+ \grad \cdot \ve Q_{K}^\win		\cr
	&=& -\left\{ \frac 1 {a\cos\lat}  
	        \bracket{\Dlon{\widemwtn {u^2}} - \widemwtn{uv} \sin\lat 
			+ \Dlat {\widemwtn{uv}\cos\lat}} 
		+ \Dp{\widemwtn{\w u}}
	    \right\} \mwtn u			\cr
	&& -\left\{ \frac 1 {a\cos\lat}
		\bracket{\Dlon {\widemwtn {uv}} + \widemwtn{u^2} \sin\lat 
		+ \Dlat {\widemwtn{v^2}\cos\lat}}
		+ \Dp {\widemwtn{\w v}}
	    \right\} \mwtn v 			\cr
	&& + \frac12 \frac 1 {a\cos\lat} 
		\Dlon\ [\widemwtn{uu} \mwtn u + \widemwtn{uv} \mwtn v] \cr
	&& + \frac12 \frac 1 {a\cos\lat} 	
		\Dlat\ [\cos\lat (\widemwtn{vu} \mwtn u + 
				\widemwtn{vv} \mwtn v)]		\cr
	&& +\frac12\Dp\ [\widemwtn{\w u} \mwtn u + \widemwtn{\w v} \mwtn v]\cr
	&=& 
	  \frac1 {2a\cos\lat} \bracket{
	        \widemwtn{u^2} \Dlon {\mwtn u} - 
		\mwtn u \Dlon {\widemwtn{u^2}}}		\cr
	&& +  \frac1 {2a\cos\lat} \bracket{
	        \widemwtn{uv} \Dlon {\mwtn v} - 
		\mwtn v \Dlon {\widemwtn{uv}}}		\cr
	&& +  \frac1 {2a\cos\lat} \bracket{
	        \widemwtn{uv} \cos\lat \Dlat {\mwtn u} - 
		\mwtn u \Dlat {\widemwtn{uv}} \cos\lat}	\cr
	&& +  \frac1 {2a\cos\lat} \bracket{
	        \widemwtn{v^2} \cos\lat \Dlat {\mwtn v} - 
		\mwtn v \Dlat {\widemwtn{v^2}} \cos\lat} \cr
	&& +  \frac12 \bracket{
		\widemwtn{\w u} \Dp {\mwtn u} - 
		\mwtn u \Dp {\widemwtn {\w u}} }	\cr
 	&& +  \frac12 \bracket{
		\widemwtn{\w v} \Dp {\mwtn v} - 
		\mwtn v \Dp {\widemwtn {\w v}} }	\cr
	&& +  \frac {\tan\lat} a \bracket{
		\mwtn u \widemwtn {uv} - \mwtn v \widemwtn{u^2}}.
	\end{eqnarray}
Obviously, the first six brackets are all in canonical form as shown in 
section~\ref{sect:separation} and hence 
represent canonical transfers. For the last term, by the property of 
marginalization (note here the dependence on $n$ is suppressed),
	\begin{eqnarray*}
	\sum_\win\sum_n 
	\bracket{\mwtn u \widemwtn {uv} - \mwtn v \widemwtn{u^2}}
	= \overline{u (uv)} - \overline{v (u^2)} = 0.
	\end{eqnarray*}
So they as a whole make $\Gamma_K^\win$ a canonical transfer.
The above formula can be further reduced to
	\begin{eqnarray}	\label{eq:Gamma_spher}
	\Gamma_K^\win 
	&=&
	  \frac1 {2a\cos\lat} \bracket{
	        \widemwtn{u^2} \Dlon {\mwtn u} - 
		\mwtn u \Dlon {\widemwtn{u^2}}
	 	+ \widemwtn{uv} \Dlon {\mwtn v} 
		- \mwtn v \Dlon {\widemwtn{uv}}
					}		\cr
	&& +  \frac1 {2a} \bracket{
	        \widemwtn{uv} \Dlat {\mwtn u} - 
		\mwtn u \Dlat {\widemwtn{uv}}
		+ \widemwtn {v^2} \Dlat {\mwtn v}
		- \mwtn v \Dlat {\widemwtn{v^2}}
					}	\cr
	&& +   \frac12 \bracket{
		\widemwtn{u \w} \Dp {\mwtn u} - \mwtn u \Dp{\widemwtn{u\w}}
	  	+ \widemwtn{v\w} \Dp{\mwtn v} - \mwtn v \Dp{\widemwtn{v\w}}
				}		\cr
	&& + \frac 3 {2a} \tan\lat \bracket{	
		\mwtn u \widemwtn{uv} - \mwtn v \widemwtn {u^2}
					}
	   + \frac 1 {2a} \mwtn v \tan\lat 
		\bracket{\widemwtn {u^2} + \widemwtn{v^2}}
	\end{eqnarray}

Note that in computing $\Gamma^\win$, we just need to perform the MWT of
nine variables, namely, the six distinct entries of the matrix
	$\left(\begin{array}{ccc}
		u^2 	&uv	&u\w	\\
		uv	&v^2	&v\w	\\
		u\w	&v\w	&\w^2
	\end{array}\right),$
plus $u$, $v$, and $\w$. The expression of $\Gamma^\win$, albeit complex,
is a combination of these variables.
The other terms can be easily expressed.

\subsection{Multiscale available potential energy equation}

Following the tradition since \cite{Lorenz}, the 
available potential energy (APE) is defined as
	\begin{eqnarray}
	A = \frac12 \frac g {\bar T (L_d-L)} T^2 \equiv \frac12 c T^2,
	\end{eqnarray}
where 
	\begin{eqnarray}
	c = \frac g {\bar T (L_d - L)} = \frac g {\bar T (g/c_p - L)}.
	\end{eqnarray}
Originally Lorenz examined the quantity in a bulk form; we relieve the
integration to define a local APE. Besides, we multiply it by $g$ to ensure
a dimension consistent with that of the kinetic energy in the preceding
section.

Multiply the heat equation (\ref{eq:atmos_gov4}) by $cT$ to get
	\begin{eqnarray*}
	\Dt A + cT \grad\cdot (\ve v T) 
		+ T\w \bar\alpha c \frac {L-L_d} g
		+ T\w\alpha c \frac {L-L_d} g 
	= cT \frac {\dot q_{net}} {c_p}.
	\end{eqnarray*}
Or
	\begin{eqnarray}
	\Dt A + \grad\cdot(\ve v A) = \alpha\w + \frac T {\bar T} \alpha\w
					+ A \w \Dp{\log c} 
					+ cT \frac{\dot q_{net}} {c_p},
	\end{eqnarray}
where $b = \alpha\w$ is the buoyancy conversion rate, 
$\frac T {\bar T} b \ll b$ is the correction term,
and $A \w \Dp{\log c}$ is the apparent source/sink due to the background
temperature profile. In the course of derivation, the ideal gas law
$\bar\alpha / \bar T =  R/p$ has been used.

To arrive at the multiscale APE equation, take an MWT on both sides
of (\ref{eq:atmos_gov4}),
followed by a multiplication with $c \mwtn T$. This gives
	\begin{eqnarray*}
	\Dt {A^\win} + c\mwtn T \grad\cdot \widemwtn{\ve v T}
	= \mwtn T \mwtn\w \frac {\bar\alpha} {\bar T}  +
	  \mwtn T \widemwtn {\w\alpha} \frac 1 {\bar T} +
	  \frac c {c_p} \mwtn T \widemwtn {\dot q_{net}}.
	\end{eqnarray*}
Write the source term as $F_A^\win$, 
	\begin{eqnarray*}
	F_A^\win = \frac c {c_p} \mwtn T \widemwtn {\dot q_{net}},
	\end{eqnarray*}
and let
	\begin{eqnarray}
	&& b^\win = \frac {\bar\alpha} {\bar T} \mwtn T \mwtn\w
		    = \frac Rp \mwtn T \mwtn\w
		    = \mwtn \alpha \mwtn\w, \\
	&& SA' = \frac 1 {\bar T} \mwtn T \widemwtn{\w\alpha}
		= \frac R p \mwtn T \widemwtn {\w T} \frac 1 {\bar T},
	\end{eqnarray}
where $b^\win$ is the buoyancy conversion rate and the other is
its correction term.
Further, separate the flux from the
transfer terms:
	\begin{eqnarray}
	&& \ve Q_{A}^\win = \frac12 c \widemwtn {\ve v T} \mwtn T	\\
	&& \Gamma_{A}^\win = -c \mwtn T \grad\cdot \widemwtn {\ve v T}
		+ \grad\cdot\ve Q_{A}^\win 
		- \frac12 \mwtn T \widemwtn {\w T} \Dp c
	\end{eqnarray}
where 
	\begin{eqnarray}
	\frac12 \mwtn T \widemwtn {\w T} \Dp c \equiv SA"
	\end{eqnarray}
is the apparent source/sink term due to the vertical variation of 
$c = c(\bar T)$.  This correction term makes $\Gamma_A^\win$ canonical. 
To see it, notice that
	\begin{eqnarray}
	\Gamma_{A}^\win 
	&& = -c \mwtn T \grad\cdot \widemwtn {\ve v T}
		+ \frac12 \grad\cdot [c\mwtn T \widemwtn {\ve vT}]
		- \frac12 \mwtn T \widemwtn {\w T}\Dp c 	\cr
	&& = \frac12c \bracket{\widemwtn{\vh T} \cdot \gradh\mwtn T 
		- \mwtn T \gradh\cdot \widemwtn{\vh T}}	 \cr
	&& \qquad
	    + \frac12 c \mwtn T \Dp{\widemwtn{\w T}}
	    + \frac12 \widemwtn{\w T} \Dp {c\mwtn T}
	    - \frac12 \mwtn T \widemwtn {\w T} \Dp c	\cr
	&&
	   = \frac c2 \bracket{\widemwtn{\vh T} \cdot \gradh\mwtn T 
		- \mwtn T \gradh\cdot \widemwtn{\vh T}}	 \cr
	&& \qquad
	    + \frac c2 \bracket{\widemwtn{\w T} \Dp {\mwtn T} 
	    -  \mwtn T \Dp {\widemwtn {\w T}}}		\cr
	&& 
	    = \frac c 2 \bracket{\widemwtn{\ve v T} \cdot \grad\mwtn T
				- \mwtn T \grad\cdot \widemwtn{\ve v T}},
	\end{eqnarray}
which is precisely in the canonical form. Following the proof in the
preceding section, it is easy to show that 
$\sum_n\sum_\win \Gamma_{A}^\win = 0$.

Combine $S'$ and $S''$ as one apparent source term to give
	\begin{eqnarray}
	S_{A}^\win = SA' + SA"
		     = \frac12 \mwtn T \widemwtn {\w T} \Dp c 
		     + \frac 1 {\bar T} \mwtn T \widemwtn{\w\alpha}.
	\end{eqnarray}
In real applications, this is usually negligible.
The multiscale APE equation now becomes
	\begin{eqnarray}
	\Dt {A^\win} + \grad\cdot\ve Q_{A}^\win
	= \Gamma_{A}^\win + b^\win + S_{A}^\win + F_A^\win.
		\label{eq:ape_atmos}
	\end{eqnarray}
In the spherical coordinates,
	\begin{eqnarray}
	&&\grad\cdot\ve Q_{A}^\win = 	
	    \frac c {2a\cos\lat} \Dlon {\bracket{\widemwtn {uT} \mwtn T}}
	  + \frac c {2a\cos\lat} \Dlat 
		{\bracket{\widemwtn {vT} \mwtn T \cos\lat}}
	  + \frac 1 2 \Dp {\bracket{c \widemwtn{\w T} \mwtn T}}	\\
	&&
	  \Gamma_{A}^\win = 
	   \frac c 2 \left[
	       \frac 1 {a\cos\lat} \widemwtn{uT} \Dlon {\mwtn T} +
	       \frac 1 a \widemwtn{vT} \Dlat {\mwtn T} +
	       \widemwtn{\w T} \Dp {\mwtn T} 	\right.		\cr
	&&\qquad
		\left.
	       - \frac 1 {a\cos\lat} \mwtn T  \Dlon {\widemwtn {uT}} 
	       - \frac 1 {a\cos\lat} \mwtn T  
			\Dlat {[\widemwtn {vT} \cos\lat]}
	       - \mwtn T \Dp {\widemwtn {\w T}}
			    \right]
	\end{eqnarray}

\subsection{A note on the units}
Currently the energetic terms have the units of $m^2/s^3$,
if the SI base units are used. However, caution should be used
when total or regional subtotal energetics are to be computed.
Since here density is not a constant,
one cannot just integrate the local fields with respect to a
volume to obtain the bulk energetics. If the system is a cartesian one,
this will be problematic, since 
	$\frac 12 \rho \vh \cdot \vh$
is NOT a quadratic variable; the variation of $\rho$ must also be taken
into account in the above derivations!

This is, however, avoidable in an isobaric frame. An integration with
respect to the ``volume'' form $dxdy (-dp)$ yields
%	\begin{eqnarray*}
%	\iiint \frac 12 \vh \cdot \vh dx dy (-dp)
%	= \iiint \frac 12 \vh \cdot \vh dxdydz \cdot \rho g
%	\end{eqnarray*}
% [in spherical coordinates this is
%	$\iiint \frac 12 \vh\cdot\vh\ a^2 \cos\lat d\lon d\lat (-dp)$],
% which is simply 
%
the real energy multiplied by a constant $g$.

\subsection{Wrap-up}
To wrap up, the multiscale kinetic and available energy equations are:
	\begin{eqnarray}
	&& \Dt {K^\win} + \grad\cdot\ve Q_K^\win = \Gamma_K^\win 
		- \grad\cdot\ve Q_P^\win - b^\win + F_{K,p}^\win +
		  F_{K,h}^\win, 	\\
	&& \Dt {A^\win} + \grad \cdot \ve Q_A^\win = \Gamma_A^\win + b^\win
		+ S_A^\win + F_A^\win.
	\end{eqnarray}
It should be mentioned that all the terms are to be multiplied by a
constant factor $2^{j_2}$, where $j_2$ is the upper bound of the scale
level of the smallest scale window. For reference, the expressions 
for the energetics are tabulated in Table~\ref{tab:atmos1}.
Also tabulated are the expressions in spherical coordinates
$(\lon,\lat,p)$ (Table~\ref{tab:atmos2}).

\begin{table}[h]
\begin{center}
\caption{Multiscale energetics for the atmospheric circulation.
	The units are in $m^2 s^{-3}$ if SI base units are used. 
	If total or regional total energies
	(in W) are to be computed, the resulting integrals with respect to
	$(x,y,p)$ should be divided by $g$.
	All terms are to be multiplied by $2^{j_2}$.  \label{tab:atmos1}}
\begin{tabular}{lll}% {p{0.75cm}  p{6.5cm}}
\hline
\hline
$K^\win$  		& $\frac12 \vhmwtn \cdot \vhmwtn$	
			   & KE on scale window $\win$ 	\\
$\ve Q_K^\win$		& $\frac12 \widemwtn{\ve v\vh} \cdot \vhmwtn$	
			   & flux of KE on window $\win$		\\
$\Gamma_K^\win$		& $\frac12 [\widemwtn{\ve v\vh} : \grad\vhmwtn
			- \grad\cdot \widemwtn{\ve v\vh} \cdot\vhmwtn]$ 	
			   & canonical transfer of KE to window $\win$	\\
$\ve Q_P^\win$ 		& $\mwtn{\ve v} \mwtn\gp$	
			   & pressure flux				\\
$b^\win$		& $\mwtn\w \mwtn\alpha$	
			   & buoyancy conversion		\\
$A^\win$		& $\frac12 c (\mwtn T)^2$,\ \ \ 
			  $c=\frac g {\bar T (g/c_p - L)}$ 
			   & APE on scale window $\win$			\\
$\ve Q_A^\win$		& $\frac12 c\mwtn T \widemwtn {\ve v T}$ 
			   & flux of APE on window $\win$		\\
$\Gamma_A^\win$		& $\frac c 2 [\widemwtn{\ve v T} \cdot \grad\mwtn T
		   	  - \mwtn T \grad\cdot \widemwtn {\ve v T}]$ 
			   & canonical transfer of APE to window $\win$ \\
$S_A^\win$		& $\frac12 \mwtn T \widemwtn {\w T} \Dp c 
			   + \frac 1 {\bar T} \widemwtn {\w\alpha}$
			   & apparent source/sink (usually negligible)	\\
\hline
\end{tabular}
\end{center}
\end{table}

\begin{table}[h]
\begin{center}
\caption{Expansion of the canonical transfers in Table~\ref{tab:atmos1}
	 in spherical coordinates.
	  \label{tab:atmos2}}
% \begin{tabular}{p{0.75cm}  p{6.5cm}}
\begin{tabular}{ll}
\hline
\hline
$\Gamma_K^\win$		& 
	$
	  \frac1 {2a\cos\lat} \bracket{
	        \widemwtn{u^2} \Dlon {\mwtn u} - 
		\mwtn u \Dlon {\widemwtn{u^2}}
	 	+ \widemwtn{uv} \Dlon {\mwtn v} 
		- \mwtn v \Dlon {\widemwtn{uv}}
					}$  	\\
\	&
	    $\qquad +  \frac1 {2a} \bracket{
	        \widemwtn{uv} \Dlat {\mwtn u} - 
		\mwtn u \Dlat {\widemwtn{uv}}
		+ \widemwtn {v^2} \Dlat {\mwtn v}
		- \mwtn v \Dlat {\widemwtn{v^2}}
					}$	\\
\	&  $\qquad + \frac12 \bracket{
		\widemwtn{u\w} \Dp{\mwtn u} - \mwtn u \Dp{\widemwtn {u\w}}
		\widemwtn{v\w} \Dp{\mwtn v} - \mwtn v \Dp{\widemwtn {v\w}}
			    }$			\\
\	&
	   $\qquad + \frac 3 {2a} \tan\lat \bracket{	
		\mwtn u \widemwtn{uv} - \mwtn v \widemwtn {u^2}
					}$	\\
\	&
	   $\qquad + \frac 1 {2a} \mwtn v \tan\lat 
		\bracket{\widemwtn {u^2} + \widemwtn{v^2}}$ \\
$\Gamma_A^\win$	&
	   $\frac c 2 \left[
	       \frac 1 {a\cos\lat} \widemwtn{uT} \Dlon {\mwtn T} +
	       \frac 1 a \widemwtn{vT} \Dlat {\mwtn T} +
	       \widemwtn{\w T} \Dp {\mwtn T} 	\right.$	\\
\	&	
	$\qquad\left.
	       - \frac 1 {a\cos\lat} \mwtn T  \Dlon {\widemwtn {uT}} 
	       - \frac 1 {a\cos\lat} \mwtn T  
			\Dlat {[\widemwtn {vT} \cos\lat]}
	       - \mwtn T \Dp {\widemwtn {\w T}}
			    \right]$				\\
\hline
\end{tabular}
\end{center}
\end{table}

%
%  Figure fig:schematic to be inserted here
%

\begin{figure*}	[t]
	\begin{center}
	\includegraphics[angle=0, width=0.75\textwidth] 
		{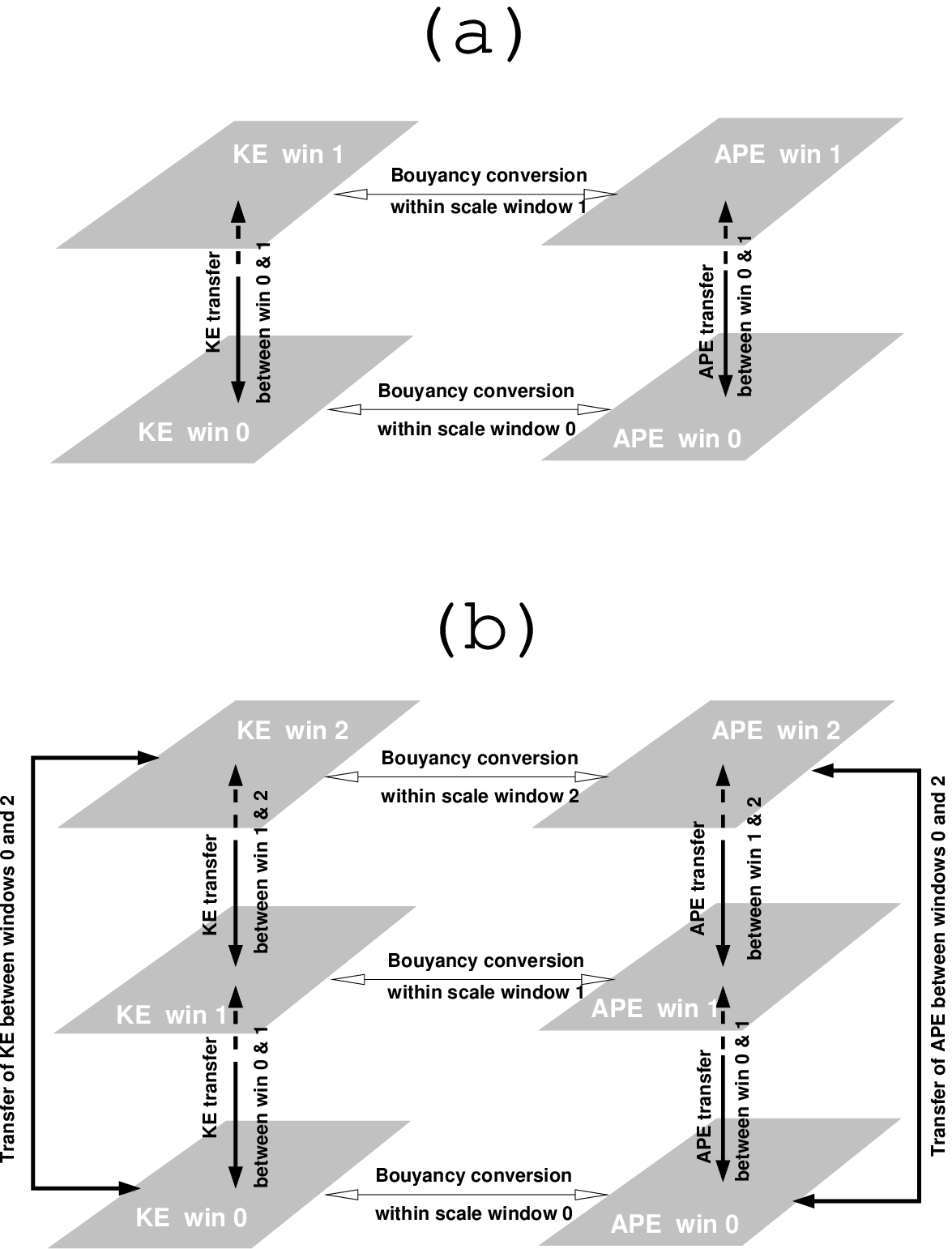}
	\caption{
	Schematic of the energy flow for (a) two-window decomposition, 
	and (b) three-window decomposition. Clearly, 
	buoyancy conversions take place within their respective scale windows;
	they are not indicators of instabilities.
	For clarity, the transfers
	$\Gamma^{0\oplus2\to1}$, $\Gamma^{1\oplus2\to0}$,
	$\Gamma^{0\oplus1\to2}$ are not drawn in (b).
	The transports of APE, KE, and pressure also take place
 	within their respective scale windows only (not shown).
	\protect{\label{fig:schematic}}}
	\end{center}
	\end{figure*}

The energy flow for a multiple-scale window decomposition 
is schematized in Fig.~\ref{fig:schematic}. 
As is seen, canonical transfers mediate between the scale windows;
they represent the interscale processes such as instabilities.
In contrast, buoyancy conversions and transports function only within
the respective individual windows; the former bring together the two types of 
energy, namely, APE and KE, while that latter allow different
spatial locations to communicate.

\section{Multiscale oceanic energetics}		\label{sect:ocean}

\subsection{Primitive equations}
The multiscale ocean energy equations have been derived in LR05.
We incorporate them here for completeness, together with
some modification and correction. 

For an incompressible and hydrostatic Boussinesq fluid flow, the primitive
equations are:
	\begin{eqnarray}
	&& \Dt \vh  + \vh \cdot \gradh \vh + w \Dz\vh + f\ve k \times \vh
	    = - \frac 1 {\rho_0} \gradh P + \ve F_{m,z} + \ve F_{m,h} 
			\label{eq:gov1_ocean}  	\\
	&& \Dz P = - \rho g	\label{eq:gov2_ocean}	\\
	&& \gradh\cdot\vh + \Dz w = 0		\label{eq:gov3_ocean}\\
	&& \Dt\rho + \vh \cdot \gradh \rho + w\Dz\rho 
	    = \frac {N^2\rho_0} g w + F_{\rho,z} + F_{\rho,h}
						\label{eq:gov4_ocean}
	\end{eqnarray}
where the subgrid process parameterization are symbolically written
as $\ve F_m$ and $F_\rho$. The other notations are referred to Appendix~A.

\subsection{Multiscale APE equation}

Following Lorenz' convention, available potential energy is defined to be
	\begin{eqnarray}
	A = \frac12 \frac {g^2} {\rho_0^2 N^2} \rho^2 
	  \equiv \frac12 c \rho^2
	\end{eqnarray}
where 
	\begin{eqnarray}
	c = \frac {g^2} {\rho_0^2 N^2} = \frac g {\rho_0 s}, \qquad
	{\rm and} \qquad s = - \Dz {\bar\rho(z)}.
	\end{eqnarray}
A recent careful discussion on Boussinesq approximation and potential
energy is referred to \cite{Ingersoll}.	% Ingersoll (2005)
As argued before, the multiscale APE on window $\win$ at step $n$ is
	$\frac12 c (\mwtn\rho )^2.$
Take MWT on both sides of the equation of density anomaly, and multiply
with $c\mwtn\rho$. It has been shown by LR05 that, to a good approximation,
$c\mwtn\rho \widemwtn {\Dt\rho}$ can be identified as $\Dt {A_n^\win}$.
The resulting APE equation is, therefore,
	\begin{eqnarray*}
	\Dt {A_n^\win} + c \mwtn\rho \widemwtn {\ve v\cdot\gradh\rho}
		       + c \mwtn\rho \widemwtn {w\Dz\rho}
	= \frac g {\rho_0} \mwtn\rho \mwtn w 
		       + F_{A,z}^\win + F_{A_h}^\win,
	\end{eqnarray*} 
where 
	\begin{eqnarray}
	\gradh\rho = \frac 1{a\cos\lat} \Dlon\rho \ve e_\lon 
		   + \frac1a \Dlat\rho \ve e_\lat,
	\end{eqnarray}
and 
	\begin{eqnarray}
	\frac g{\rho_0} \mwtn\rho \mwtn w \equiv b_n^\win
	\end{eqnarray}
is the rate of buoyancy conversion.

The key to the multiscale energetics formalism
is the separation of flux and transfer processes. 
By that in subsection~\ref{sect:separation}\ref{sect:flux}, 
the flux of APE by $\ve v$ at step $n$ on window $\win$ is
	\begin{eqnarray}
	\ve Q_{A}^\win = \frac12 c \mwtn\rho \widemwtn{\ve v\rho}.
	\end{eqnarray}
Hence the above equation can be written as
	\begin{eqnarray*}
	\Dt {A^\win} + \grad\cdot \ve Q_{A}^\win
	= \bracket{-c\mwtn\rho \widemwtn {\ve v\cdot\grad\rho} 
		   + \grad\cdot\ve Q_{A}^\win}
	  + b_n^\win  + F_{A,z}^\win + F_{A,h}^\win.
	\end{eqnarray*}
But the bracket on the r.h.s. is still not the canonical transfer
that we are seeking for. Since $c = c(z)$, it does not summarize to zero.
In fact, 
	\begin{eqnarray*}
	&& \sum_\win\sum_{n} 
		\bracket{-c\mwtn\rho \widemwtn {\ve v \cdot \grad\rho} 
	        + \grad\cdot\ve Q_{A}^\win}
	= -c \overline {\rho\grad\cdot(\rho\ve v)} 
	  + \frac12\grad\cdot \bracket{c \overline{\rho(\ve v\rho)}} \cr
	&&\qquad = \frac12 \Dz c \overline{\rho^2 w},
	\end{eqnarray*}
where the overbar denotes averaging over the time period. Write
	\begin{eqnarray}
	S_{A,n}^\win = \frac12 \Dz c \mwtn\rho \widemwtn {\rho w}
	\end{eqnarray}
% (or $\frac12 \Dz c \widemwtn{\rho^2} \mwtn w $?), 
which is the apparent
source/sink due to the vertical stratification. Then 
	\begin{eqnarray}	\label{eq:TA}
	\Gamma_{A}^\win 
	&=& 
	  \bracket{-c\mwtn\rho \widemwtn {\ve v \cdot \grad\rho} 
	        + \grad\cdot\ve Q_{A,n}^\win} - S_{A,n}^\win	\cr
	&=&
	  \frac c 2 \bracket{\widemwtn {\ve v\rho} \cdot \grad\mwtn\rho
			- \mwtn\rho \grad\cdot \widemwtn {\ve v\rho}}
	\end{eqnarray}
proves to be canonical.
The multiscale APE equation is, accordingly,
	\begin{eqnarray}
	\Dt {A^\win} + \grad\cdot \ve Q_{A}^\win
	= \Gamma_{A}^\win
	  + b^\win  + S_{A}^\win + F_{A,z}^\win + F_{A,h}^\win.
	\end{eqnarray}

In the spherical coordinate frame, 
% the Lam\'e's coefficients are
%	$h_\lon \approx a\cos\lat,
%	 h_\lat \approx a,
%	 h_z = 1,$
% thus the divergence of, say, $\ve Q = (Q_\lon, Q_\lat, Q_z)$, is
%	\begin{eqnarray}
%	\grad\cdot\ve Q 
%	&=& \frac 1 {h_\lon h_\lat h_z} \bracket{
%		\Dlon {(h_\lat h_z Q_\lon)} + \Dlat {(h_\lon h_z Q_\lat)} 
%		+ \Dz{(h_\lon h_\lat Q_z)} 
%						      } \cr
%	&=& \frac 1 {a\cos\lat} \Dlon {Q_\lon} +
%	    \frac 1 {a\cos\lat} \Dlat {(Q_\lat \cos\lat)} + 
%	    \Dz {Q_z}.
%	\end{eqnarray}
by (\ref{eq:divQ}), we have
	\begin{eqnarray}
	\grad\cdot\ve Q_A^\win 
	&=& \frac c {2a\cos\lat} \bracket{
		\Dlon\ \parenth{\mwtn\rho \widemwtn {u\rho}} +
		\Dlat\ \parenth{\mwtn\rho \widemwtn{v\rho} \cos\lat}
					}
	   + \frac12 \Dz\ \bracket{c \mwtn\rho \widemwtn{w\rho}},
	\end{eqnarray}
and
	\begin{eqnarray}
	\Gamma_A^\win 
	&& = \frac c 2 \left[
	 \frac1 {a\cos\lat} \widemwtn{u\rho} \Dlon {\mwtn\rho}
	+ \frac 1a \widemwtn{v\rho} \Dlat {\mwtn\rho}
	+ \widemwtn{w\rho} \Dz{\mwtn\rho}  \right.		\cr
	&&\ \ 
	  \left.
  	   - \mwtn\rho \parenth{
		\frac 1 {a\cos\lat} \Dlon {\widemwtn{u\rho} }
		+ \frac 1 {a\cos\lat} \Dlat {\widemwtn{v\rho} \cos\lat}
		+ \Dz {\widemwtn{w\rho}}
			       }
					  \right]
	\end{eqnarray}

\subsection{The KE equation}

The equation governing the evolution of the multiscale kinetic energy (KE)
	\begin{eqnarray}
	K^\win = \frac12 \mwtn{\ve v} \cdot \mwtn{\ve v}
	\end{eqnarray}
can be obtained by taking MWT on both sides of
the horizontal momentum equations, followed by a dot product with
$\mwtn{\ve v}$. This results in
	\begin{eqnarray*}
	&& \Dt {K_n^\win} + \widemwtn{\ve v\cdot\grad\vh} \cdot\vhmwtn
	= -\frac1{\rho_0} \vh\cdot \gradh\mwtn P 
		+ F_{K,z}^\win + F_{K,h}^\win\\
	&&
	= - \ve v \cdot \grad \frac{\mwtn P} {\rho_0} 
		+ \mwtn w \Dz {\mwtn P / \rho_0}
		+ F_{K,z}^\win + F_{K,h}^\win\\
	&&
	= - \grad \cdot \parenth{\mwtn{\ve v} \frac{\mwtn P} {\rho_0}}
		- \frac g {\rho_0} \mwtn\rho \mwtn w
		+ F_{K,z}^\win + F_{K,h}^\win	\\
	&&
	\equiv - \grad\cdot \ve Q_{P}^\win - b^\win
		+ F_{K,z}^\win + F_{K,h}^\win,
	\end{eqnarray*}
where $b^\win$ is the buoyancy conversion rate, and 
$-\grad\cdot \ve Q_{P}^\win$ the pressure working rate.
In the above derivation 
Eqs.~(\ref{eq:gov2_ocean}) and (\ref{eq:gov3_ocean}) (incompressibility and
hydrostaticity) have been used.

By the transport-transfer separation, the above
multiscale KE equation can be written as
	\begin{eqnarray}
	\Dt {K^\win} + \grad\cdot\ve Q_{K}^\win
	= 
		\Gamma_K^\win
	       - \grad\cdot \ve Q_{P}^\win - b^\win
		+ F_{K,z}^\win + F_{K,h}^\win,
	\end{eqnarray}
where 
	\begin{eqnarray}
	\ve Q_K^\win = \frac12 \widemwtn{\ve v \vh} \cdot \vhmwtn,
	\end{eqnarray}
and
	\begin{eqnarray}
	\Gamma_K^\win
	&=& - \widemwtn{\ve v\cdot\grad\vh} \cdot\vhmwtn 
	    + \grad\cdot\ve Q_{K}^\win	\cr
	&=& - \grad\cdot\widemwtn{\ve v\vh} 
		\cdot \vhmwtn + \grad\cdot\ve Q_{K,n}^\win	\cr
	&=&
	 \frac12 \bracket{\widemwtn {\ve v \vh} : \grad\vhmwtn
		- [\grad\cdot\widemwtn {\ve v\vh}] \cdot \vhmwtn},
	\end{eqnarray}
which are precisely the same as that for the atmosphere case.
In the spherical coordinates $(\lon, \lat, z)$, 
$\grad\cdot\ve Q_K^\win$ and $\Gamma_K^\win$ are also like that
in (\ref{eq:Q_spher}) and (\ref{eq:Gamma_spher}), 
except that $\w$ should be replaced by $w$, and $p$ by $z$.

\subsubsection{Wrap-up}
To wrap up, the multiscale ocean energetic equations are
	\begin{eqnarray}
	&& \Dt {A^\win} + \grad\cdot Q_{A}^\win
	= \Gamma_{A}^\win + b^\win + S_{A}^\win + 
	   F_{A,z}^\win + F_{A,h}^\win,	\label{eq:ke_ocean} \\
	&& \Dt {K^\win} + \grad\cdot Q_{K}^\win
	= \Gamma_{K^\win} - \grad\cdot\ve Q_{P}^\win - b^\win +
	   F_{K,z}^\win + F_{K,h}^\win.	\label{eq:ape_ocean}
	\end{eqnarray}
The expressions are referred to Table~\ref{tab:ocean1}.

\begin{table}[h]
\begin{center}
\caption{Multiscale energetics for oceanic circulations.
	The expressions in spherical coordinates are the same
	in form as that in Table~\ref{tab:atmos2}, except that
	the coordinate $p$ should be replaced by $z$ and $T$ by $\rho$.
	The units are in $m^2 s^{-3}$ (if SI units are used). 
	All terms are to be multiplied by $2^{j_2}$.  \label{tab:ocean1}}
\begin{tabular}{lll}  %p{0.75cm}  p{6.5cm}}
\hline
\hline
$K^\win$  		& $\frac12 \vhmwtn \cdot \vhmwtn$	
			   &	KE on scale window $\win$ 	\\
$\ve Q_K^\win$		& $\frac12 [\widemwtn{\ve v\vh} \cdot \vhmwtn]$
			   & flux of KE on window $\win$	\\
$\Gamma_K^\win$		& $\frac12 [\widemwtn{\ve v\vh} : \grad\vhmwtn
			- \grad\cdot \widemwtn{\ve v\vh} \cdot\vhmwtn]$  
			   & canonical transfer of KE to window $\win$ \\
$\ve Q_P^\win$ 		& ${\frac 1 {\rho_0} \mwtn{\ve v} \mwtn P}$
			   & pressure flux on window $\win$	\\
$b^\win$		& $\frac g {\rho_0} \mwtn\rho \mwtn w$	
			   & buoyancy conversion on window $\win$ \\
$A^\win$		& $\frac12 c (\mwtn \rho)^2$,\ \ \ 
			  $c=\frac {g^2} {\rho_0^2 N^2}$ 	
			   & APE on window $\win$ \\
$\ve Q_A^\win$
			& $\frac12 [c\mwtn \rho \widemwtn {\ve v \rho}]$ 
			   & flux of APE on window $\win$	\\
$\Gamma_A^\win$		& $\frac c 2 [\widemwtn{\ve v \rho} 
			   \cdot \grad\mwtn \rho - \mwtn \rho \grad
			   \cdot \widemwtn {\ve v \rho}]$ 
			   & canonical transfer of APE to window $\win$ \\
$S_A^\win$		& $\frac12 \mwtn \rho \widemwtn {\w \rho} \Dz c$
			   & apparent source/sink of $A^\win$
			   (usually negligible)			\\
\hline
\end{tabular}
\end{center}
\end{table}

\section{Multiscale quasi-geostrophic energetics}	\label{sect:QG}

The multiscale energy equations like (\ref{eq:ke_atmos}) and
(\ref{eq:ape_atmos}) cannot be directly derived from the quasi-geostrophic
(QG) equation. We have to go back to where the QG equation comes from and 
do the derivation, and this is how 
	 \cite{Pinardi}
	% Pinardi and Robinson (1986) 
did with their regional QG energetics.

Since the atmospheres and oceans share the same QG equation, it suffices
to start off the derivation from either
(\ref{eq:atmos_gov1})-(\ref{eq:atmos_gov5})
or (\ref{eq:gov1_ocean})-(\ref{eq:gov4_ocean}).
As vertically a $z$ coordinate is desired, we choose the latter.
To simplify the presentation, the dissipative and diffusive processes are
omitted. They are not essential to the derivation, and their effect may be
added symbolically after the other terms are finalized.
From Appendix~C, the QG equation we will be dealing with is:
	\begin{eqnarray}		\label{eq:QG_gov}
        \Dt\ \bracket{\gradh^2\psi + 
		\Dz\ \parenth{\frac{\Fr^2}{N^2}\Dz\psi}}  +
         \nlcoef J\parenth{\psi,\ \bracket{\gradh^2\psi + 
		\Dz\ \parenth{\frac{\Fr^2}{N^2}\Dz\psi}}
	  } 
	 + \beta \Dx\psi = 0,
	\end{eqnarray}
% [cf.~Eq.~(\ref{app:qg8})] are
where $\Fr$ is the rotational internal Froude number,
$\nlcoef$ a dimensionless measure of the importance of advection,
and $J$ the Jacobian operator; the other notations are conventional and
referred to Appendix~A.

\subsection{QG kinetic energetics}
The inviscid version of the KE equation (\ref{eq:ke_ocean}) is rewritten as
	\begin{eqnarray}	\label{eq:ke_ocean2}
	\Dt {K^\win} + \gradh\cdot {\ve Q}_{K,h}^\win + \Dz {Q_{K,z}^\win}
		= \Gamma_{K,h}^\win + \Gamma_{K,z}^\win
		  - \gradh\cdot {\ve Q}_{P,h}^\win - \Dz {Q_{P,z}^\win}
		  - b^\win, 
	\end{eqnarray}
where
	\begin{eqnarray}
	&& \Gamma_{K,h}^\win = -\gradh \cdot \widemwtn{\vh\vh} \cdot \vhmwtn
			    + \gradh\cdot\ve Q_{K,h}^\win, \\
	&& \Gamma_{K,z}^\win = -\Dz {\widemwtn {w\vh}} \cdot \vhmwtn 
			    + \Dz{Q_{K,z}^\win}.
	\end{eqnarray}
In the transport and transfer terms, the effects due to the horizontal and
vertical advections are distinguished. 
As we will see soon, this will greatly help simplify the QG energetics.

Using the usual scaling (e.g., \cite{McWilliams_book}),
		% (e.g., McWilliams, 2006),
	\begin{eqnarray*}
	&& (x,y) \sim L_0, \qquad z\sim H_0, \qquad t\sim t_0, \\
	&& (u,v) \sim U_0, \qquad w \sim \frac {H_0} {L_0} U_0, \\
	&& f\sim f_0, \qquad N\sim N_0, \\
	&& P\sim U_0 f_0 \rho_0 L_0, \qquad
	   \rho \sim \frac{f_0 U_0 L_0} {gH_0} \rho_0,
	\end{eqnarray*}
and noticing that the multiscale window transform does not affect the scaling,
it is easy to have
	\begin{eqnarray*}
	&& \Dt{K^\win} \sim \frac {U_0^2} {t_0}; \\
	&& \gradh\ve Q_{K,h}^\win\ {\rm and}\ \Dz{Q_{K,z}^\win}
	    \ \sim\ \frac{U_0^3} {L_0}; 			\\
	&& \Gamma_{K,h}^\win\ {\rm and}\ \Gamma_{K,z}^\win
	    \ \sim\ \frac{U_0^3} {L_0};	\\
	&& \gradh\cdot\ve Q_{P,h}^\win\ {\rm and}\ \Dz{Q_{P,z}^\win}
	    \ \sim U_0^2 f_0; 	\\
	&& b^\win = \frac g {\rho_0} \mwtn\rho \mwtn w \sim U_0^2 f_0.
	\end{eqnarray*}
This will yield the nondimensionalized kinetic energetics.

For clarity, {\it hereafter throughout this subsection, all
variables are understood as nondimensional}. From above, 
Eq.~(\ref{eq:ke_ocean2}) is now reduced to its nondimensional form:
	\begin{eqnarray}	\label{eq:ke_qg_tmp}
	\eps \Dt {K^\win} + \eps\nlcoef \gradh\cdot {\ve Q}_{K,h}^\win 
	+ \eps\nlcoef \Dz {Q_{K,h}^\win}
	= \eps\nlcoef \Gamma_{K,h}^\win + \eps\nlcoef \Gamma_{K,z}^\win
	  - \gradh\cdot {\ve Q}_{P,h}^\win - \Dz {Q_{P,z}^\win}
		  - b^\win, 
	\end{eqnarray}
where $\eps = \frac 1 {f_0 t_0}$ is the Rossby number, 
$\nlcoef = \frac {U_0 t_0} {L_0}$ measures the relative importance
of advection to local change. In many textbooks, $\nlcoef$ is taken to be
one, so that $\eps\nlcoef = U_0/f_0L_0$ is defined as the Rossby number.

As usual, expand the variables in the power of $\eps$, 
	\begin{eqnarray}
	&& P = [P]_0 + \eps [P]_1 + \eps^2 [P]_2... \label{eq:qg_epsP}	\\
	&& w = [w]_0 + \eps [w]_1 + \eps^2 [w]_2...	\\
	&& \vh = [\vh]_0 + \eps [\vh]_1 + \eps^2 [\vh]_2...\\
	&& \rho = [\rho]_0 + \eps [\rho]_1 + \eps^2 [\rho]_2... 
						\label{eq:qg_epsrho}
	\end{eqnarray}
Based on these expansions the multiscale energetic terms can also be 
expanded. For example, 
	\begin{eqnarray*}
	K = [K]_0 + \eps [K]_1 + \hdots
	\end{eqnarray*}
where $[K]_0 = \frac12 [\vh]_0 \cdot [\vh]_0$,
      $[K]_1 = [\vh]_0 \cdot [\vh]_1$, 
and so forth.
By the classical result [see (\ref{app:qg1})-(\ref{app:qg3}) 
in App.~C], $[w]_0 = 0$.
So 
	\begin{eqnarray*}
	\eps \Dz{Q_{K,z}^\win} = \eps \Dz\ \frac12
		\cbrace{\widemwtn {w\vh} \cdot {\vh}}\ \sim\ O(\eps^2).
	\end{eqnarray*}
Likewise,
	\begin{eqnarray*}
	&& \eps\Gamma_{K,z}^\win \sim O(\eps^2), \\
	&& \Dz\ Q_{P,z}^\win = \Dz\ (\mwtn w \mwtn P)
		= \eps \Dz\ ([\mwtn w]_1 [\mwtn P]_0) + O(\eps^2), \\
	&& b^\win = \mwtn\rho \mwtn w 
		= \eps [\mwtn\rho]_0 [\mwtn w]_1 + O(\eps^2).
	\end{eqnarray*}
Substituting the power expansions into (\ref{eq:ke_qg_tmp}), taking
into account the above facts, and equating the terms of like power,
we have, to the order of $O(\eps^0)$,
	\begin{eqnarray*}
	\qquad \gradh\cdot \parenth{[\vhmwtn]_0 [\mwtn P]_0} = 0.
	\end{eqnarray*}
So a huge part of the pressure working rate is actually zero. 
To the order of $O(\eps^1)$,
	\begin{eqnarray*}
	&&\Dt{[K^\win]_0} + \nlcoef \gradh\cdot [\ve Q_{K,h}^\win]_0
	= \nlcoef [\Gamma_{K,h}^\win]_0
    	  - \gradh \cdot [\ve Q_{P,h}^\win]_1 - \Dz\ [Q_{P,z}^\win]_1
	  - [b^\win]_1    \\
	&&
	= \nlcoef \Gamma_{K,h}^\win 
	  - \gradh \cdot \parenth{[\vhmwtn]_0 [\mwtn P]_1 
			        + [\vhmwtn]_1 [\mwtn P]_0}
	  - \Dz\ \parenth{[\mwtn w]_1 [\mwtn P]_0}
	  - [\mwtn\rho]_0 [\mwtn w]_1.
	\end{eqnarray*}

To this order, $[\vh]_0$ is the geostrophic flow: 
$[\vh]_0 = \ve k \times \grad [P]_0$, 
$[\rho]_0$ is $-\Dz\ [P]_0$ by hydrostaticity.
$[w]_1$ and $[\vh]_1$ can also be obtained 
[see (\ref{app:qg6})-(\ref{app:qg7}) in App.~C]:
	\begin{eqnarray*}
	&& [w]_1 = -\frac {\Fr^2} {N^2} {\mathscr L} 
				\parenth{\Dz {[P]_0}}, \\
	&& [\vh]_1 = \ve k \times {\mathscr L}([\vh]_0) - \beta y[\vh]_0
			+ \ve k \times \gradh[P]_1,
	\end{eqnarray*}
where $\Fr = \frac{f_0L_0} {N_0H_0}$ is the rotational internal Froude
number, and  $\mathscr L$ stands for the operator
	\begin{eqnarray}
	{\mathscr L} \equiv \Dt\ + \nlcoef [\vh]_0 \cdot \gradh
	\end{eqnarray}
(advection by the geostrophic flow).
With these, it is straightforward to compute
$[K^\win]_0$, $[\ve Q_{K,h}^\win]_0$, $[\Gamma_{K,h}^\win]_0$,
$\Dz\ [Q_{P,z}^\win]_1$ and $[b^\win]_1$.
The difficulty comes from the horizontal pressure working rate
$[\ve Q_{P,h}^\win]_1$, where $[\mwtn P]_1$ is involved.
But
	\begin{eqnarray*}
	&&
	\gradh \cdot \parenth{[\vhmwtn]_0 [\mwtn P]_1 
			        + [\vhmwtn]_1 [\mwtn P]_0} 	\\
	&&= 
	\gradh \cdot \parenth{[\vhmwtn]_0 [\mwtn P]_1 
	+ (\ve k\times \widemwtn {{\mathscr L}[\vh]_0} - \beta y [\vhmwtn]_0
			+ \ve k \times \gradh[\mwtn P]_1) [\mwtn P]_0}  \\
	&&=
	\gradh \cdot ((\ve k \times \widemwtn {{\mathscr L}[\vh]_0} 
				- \beta y [\vhmwtn]_0) [\mwtn P]_0)
	+ \gradh \cdot ([\vhmwtn]_0 [\mwtn P]_1
			+ \ve k \times \gradh[\mwtn P]_1 [\mwtn P]_0).
	\end{eqnarray*}
Notice that the second divergence vanishes. In fact, it is
	\begin{eqnarray*}
	&& \gradh \cdot (\ve k \times \gradh[\mwtn P]_0 [\mwtn P]_1
			+ \ve k \times \gradh[\mwtn P]_1 [\mwtn P]_0) \\
	&& = -\ve k \cdot \gradh \times 
        (\gradh [\mwtn P]_0 [\mwtn P]_1 + \gradh [\mwtn P]_1 [\mwtn P]_0) \\
	&& = 0.
	\end{eqnarray*}
Hence the whole pressure working rate
	\begin{eqnarray}
	\gradh \cdot [\ve Q_{P,h}^\win]_1 
	= \gradh \cdot ((\ve k \times \widemwtn {{\mathscr L}[\vh]_0} 
				- \beta y[\vhmwtn]_0) [\mwtn P]_0).
	\end{eqnarray}

As a convention, denote $[P]_0$ as $\psi$, and for convenience, 
write $[\vh]_0 = \ve k \times \gradh\psi$ as $\ve v_{g}$ (geostrophic
velocity). Distinguishing the QG energetics terms with a subscript $g$,
the multiscale KE now becomes
	\begin{eqnarray}
	\Dt\ K_{g}^\win + \gradh\cdot \ve Q_{K,g}^\win
	= \Gamma_{K,g}^\win - \gradh\cdot \ve Q_{P,g,h}^\win
	  - \Dz\ Q_{P,g,z}^\win - b_{g}^\win,
	\end{eqnarray}
where
	\begin{eqnarray}
	&& K_{g}^\win = \frac12 \mwtn {\ve v_{g}} \cdot \mwtn {\ve v_{g}},\\
	&& Q_{K,g}^\win = \nlcoef \ve [Q_{K,h}^\win]_0 = 
	   \frac12 \nlcoef \widemwtn {\ve v_{g} \ve v_{g}} 
		\cdot \mwtn{\ve v_{g}}, 			\\
	&& \Gamma_{K,g}^\win = \nlcoef [\Gamma_{K,h}^\win]_0 = 
	   \frac \nlcoef 2 \bracket{\widemwtn{\ve v_{g} \ve v_{g}} : 
			\gradh \mwtn {\ve v_{g}}
		- \gradh \cdot \widemwtn{\ve v_{g} \ve v_{g}}
			\cdot \mwtn{\ve v_{g}} 	      } \\
	&& \ve Q_{P,g,h}^\win = (\ve k \times 
		{\mathscr L}(\mwtn{\ve v_{g}}) - \beta y\mwtn{\ve v_{g}}) 
		\mwtn\psi	\\
	&& Q_{P,g,z}^\win = - \frac{\Fr^2}{N^2} \mwtn\psi
		\widemwtn {{\mathscr L}\parenth{\Dz\psi}} 	\\
	&& b_{g}^\win = \frac{\Fr^2} {N^2} \Dz {\mwtn\psi} 
		\widemwtn {{\mathscr L}\parenth{\Dz\psi}} 
	\end{eqnarray}
(recall that all are to be multiplied by a constant fact $2^{j_2}$).

%%%%%%%%%%%%%%%%%%%%%%%%%%%%%%%%%%%%%%%%%%%%%%%%%%%%%%%%%%%%%%%%%%%%

\subsection{QG available potential energetics}
Rewrite the nondiffusive version of the APE equation (\ref{eq:ape_ocean}) 
as
	\begin{eqnarray}	\label{eq:ape_ocean2}
	\Dt {A^\win} + \gradh\cdot {\ve Q}_{A,h}^\win + \Dz {Q_{A,z}^\win}
		= \Gamma_{A,h}^\win + \Gamma_{A,z}^\win + b^\win
		  + S_A^\win.
	\end{eqnarray}
Using the scaling as shown in the preceding subsection, we have
	\begin{eqnarray*}
	&& \Dt {A^\win} = \frac12 \frac {g^2} {\rho_0^2 N^2} 
		       \parenth{\mwtn\rho}^2
		     \sim \frac 1 {t_0} \frac {g^2} {\rho_0^2 N_0^2} 
		 \parenth{\frac {f_0U_0L_0} {gH_0} \rho_0}^2	\cr
	&&\qquad
		 = \frac {U_0^2} {t_0} \frac {f_0^2L_0^2}{N_0^2H_0^2}
		 \equiv \frac {U_0^2} {t_0} \Fr^2	\\
	&& \gradh \cdot \ve Q_{A,h}^\win 
		\sim \frac 1 {L_0} \frac {g^2} {\rho_0^2N_0^2} U_0
		     \parenth{\frac {f_0U_0L_0} {gH_0} \rho_0}^2
	     	= \frac {U_0^3} {L_0} \Fr^2,
	\end{eqnarray*}
where $\Fr = \frac{f_0 L_0} {N_0 H_0}$ is the rotational internal Froude
number (compared to the Froude number $\frac {U_0} {N_0 H_0}$).
Likewise, all the remaining terms, except $b^\win \sim U_0^2f_0$ (as given
in the preceding subsection), 
are of the order of $\frac{U_0^3} {L_0} \Fr^2$.

As in the preceding subsection,
let $\eps = \frac 1 {f_0 t_0}$ be the Rossby number and let 
$\nlcoef = \frac {U_0 t_)} {L_0}$. 
The scaled nondiffusive APE equation 
({\it from now on throughout this subsection all the
variables are nondimensional}) is, therefore, 
	\begin{eqnarray}
	\eps\Fr^2 \parenth{\Dt {A^\win} + 
	  \nlcoef \gradh\cdot\ve Q_{A,h}^\win  + \nlcoef \Dz\ Q_{A,z}^\win}
	= \eps\Fr^2\nlcoef (\Gamma_{A,h}^\win + \Gamma_{A,z}^\win)
	  + b^\win + \eps\Fr^2 S_A^\win.
	\end{eqnarray}
Usually $\Fr$ is taken as order of $O(1)$, but $\eps$ is small. 
Expanding in the power of $\eps$, since $[w]_0 = 0$ (cf.~App.~C),
it is easy to show that
	\begin{eqnarray*}
%	&& b^\win \sim O(\eps), \\
	\eps\Dz\ Q_{A,z}^\win \sim \eps \Gamma_{A,z}^\win \sim 
		\eps S_A^\win\ \sim\ O(\eps^2).
	\end{eqnarray*}
In other words, when only order of $O(\eps)$ is considered, all these terms
are negligible. Therefore, the resulting APE equation is, to the order of
$O(\eps)$,
	\begin{eqnarray}
	\Fr^2 \Dt {[A^\win]_0} + \nlcoef\Fr^2 \gradh\cdot [\ve Q_{A,h}^\win]_0
		= \nlcoef\Fr^2 [\Gamma_{A,h}^\win]_0 + [b^\win]_1.
	\end{eqnarray}
For clarity, this is symbolically written as
	\begin{eqnarray}
	\Dt\ A_{g}^\win + \gradh\cdot \ve Q_{g,A}^\win
		= \Gamma_{g,A}^\win + b_{g}^\win,
	\end{eqnarray}
where
	\begin{eqnarray*}
	&& A_{g}^\win = \Fr^2 [A^\win]_0 = 
		\frac {\Fr^2} 2 \frac 1 {N^2} \parenth{\Dz {\mwtn\Psi}}^2 \\
	&& \ve Q_{g,A}^\win = \nlcoef\Fr^2 [\ve Q_{A,h}^\win]_0 = 
		\frac\nlcoef 2 \frac{\Fr^2} {N^^2} 
		\Dz {\mwtn\Psi} \widemwtn{\ve v_{g} \Dz\Psi}   	 \\
	&& \Gamma_{g,A}^\win = \nlcoef\Fr^2 [\Gamma_{A,h}^\win]_0 = 
		\frac\nlcoef 2 \frac{\Fr^2} {N^2} 
		\bracket{
		   \widemwtn{\ve v_{g} \Dz\Psi} \cdot \gradh\Dz{\mwtn\Psi} 
	 	   - \Dz{\mwtn\Psi} \gradh\cdot \widemwtn{\ve v_{g} \Dz\Psi}
		        } 				\\
	&& b_{g}^\win = [b^\win]_1 = 
		\frac{\Fr^2} {N^2} \Dz {\mwtn\psi} 
		\widemwtn {{\mathscr L}\parenth{\Dz\psi}} 
	\end{eqnarray*}

\subsection{Wrap-up}
To summarize, the multiscale energy equations for the inviscid QG equation
(\ref{eq:QG_gov}) are
% 	\begin{eqnarray}		\label{eq:QG_gov}
%        \Dt\ \bracket{\gradh^2\psi + 
%		\Dz\ \parenth{\frac{\Fr^2}{N^2}\Dz\psi}}  +
%         \nlcoef J\parenth{\psi,\ \bracket{\gradh^2\psi + 
%		\Dz\ \parenth{\frac{\Fr^2}{N^2}\Dz\psi}}
%	  } 
%	 + \beta \Dx\psi = 0
%	\end{eqnarray}
% [cf.~Eq.~(\ref{app:qg8})] 
%
	\begin{eqnarray}
	&& \Dt\ A_{g}^\win + \gradh\cdot \ve Q_{g,A}^\win
		= \Gamma_{g,A}^\win + b_{g}^\win,	 \\
	&& \Dt\ K_{g}^\win + \gradh\cdot \ve Q_{g,K}^\win
	= \Gamma_{g,K}^\win - \gradh\cdot \ve Q_{g,P,h}^\win
	  - \Dz\ Q_{g,P,z}^\win - b_{g}^\win.
	\end{eqnarray}
The explicit expressions of the energetic terms are tabulated in 
Table~\ref{tab:qg1}.

\begin{table}[h]
\begin{center}
\caption{Expansion of the QG energetics for Eq.~(\ref{eq:QG_gov}). 
	$\ve v_g = \ve k \times \gradh\psi$, 
	${\mathscr L} = \Dt\ + J(\psi, .)$.
	  \label{tab:qg1}}
\begin{tabular}{lll}		% {p{0.75cm}  p{6.5cm}}
\hline
\hline
$K_{g}^\win$ 	      & $\frac12 \mwtn {\ve v_{g}} \cdot \mwtn {\ve v_{g}}$
			 & QG KE on scale window $\win$ 	\\
$Q_{K,g}^\win$        & $\frac12 \nlcoef \widemwtn {\ve v_{g} \ve v_{g}} 
		       \cdot \mwtn{\ve v_{g}}$ 
			 & Flux of QG KE within scale window $\win$	\\
$\Gamma_{K,g}^\win$   & $\frac \nlcoef 2 \bracket{
			\widemwtn{\ve v_{g} \ve v_{g}} : 
			\gradh \mwtn {\ve v_{g}}
		         - \gradh \cdot \widemwtn{\ve v_{g} \ve v_{g}}
			\cdot \mwtn{\ve v_{g}}  }$
			& canonical transfer of QG KE to window $\win$	\\
$\ve Q_{P,g,h}^\win$  &  $[\ve k \times 
   \widemwtn{{\mathscr L}(\ve v_{g})} - \beta y \mwtn{\ve v_{g}}] \mwtn\psi$
			& horizontal pressure flux on window $\win$	\\
$Q_{P,g,z}^\win$      &  $-\frac{\Fr^2}{N^2} \mwtn\psi \bracket{
				\frac{\D^2\mwtn\psi} {\D t\D z}   
			    + \nlcoef \widemwtn {\ve v_g \cdot\gradh \Dz\psi}
			   				       }$
			& vertical pressure flux on window $\win$	\\
$b_{g}^\win$            &  $\frac{\Fr^2} {N^2} \Dz {\mwtn\psi} 
		          \widemwtn {{\mathscr L}\parenth{\Dz\psi}}$
			   & rate of buoyancy conversion on window $\win$ \\
$A_{g}^\win$ 		& $\frac {\Fr^2} 2 \frac 1 {N^2} 
			  \parenth{\Dz {\mwtn\Psi}}^2$ 	
			   & QG APE on scale window $\win$ 		\\
$\ve Q_{A,g}^\win$ 	& $ \frac\nlcoef 2 \frac{\Fr^2} {N^^2} 
			   \Dz {\mwtn\Psi} \widemwtn{\ve v_{g} \Dz\Psi}$
			   & flux of QG APE within window $\win$ 	\\
$\Gamma_{A,g}^\win$ 	& $ \frac\nlcoef 2 \frac{\Fr^2} {N^2} 
			   \bracket{
		   		\widemwtn{\ve v_{g} \Dz\Psi} \cdot 
					\gradh\Dz{\mwtn\Psi} 
	 	   		- \Dz{\mwtn\Psi} \gradh\cdot 
				\widemwtn{\ve v_{g} \Dz\Psi}
		        	   }$
			   & canonical transfer of QG APE to window $\win$\\
\hline
\end{tabular}
\end{center}
\end{table}

\begin{table}
 \begin{center}
 \caption{Expansion of the QG canonical transfers 
 	 in spherical coordinates.  \label{tab:qg2}}
\begin{tabular}{ll}
 \hline
 \hline	
 $\Gamma_{K,g}$	   & $\frac\nlcoef {2a\cos\lat}\bracket{
		   \widemwtn{u_g^2} \Dlon{\mwtn u_g}
		   - \mwtn u_g \Dlon {\widemwtn{u_g^2}}
		   + \widemwtn{u_gv_g} \Dlon {\mwtn v_g}
		   - \mwtn v_g \Dlon {\widemwtn {u_gv_g}}
						    }$ \\
 \		& $+ \frac\nlcoef {2a} \bracket{
		   \widemwtn{u_g v_g} \Dlat{\mwtn u_g}
		   - \mwtn u_g \Dlat {\widemwtn {u_g v_g}}
		   + \widemwtn {v_g^2} \Dlat {\mwtn v_g}
		   - \mwtn v_g \Dlat {\widemwtn {v_g^2}}
						       }$	\\
 \		& $+ \frac{3\nlcoef} {2a} \tan\lat \bracket{
		  \mwtn u_g \widemwtn {u_g v_g} - \mwtn v_g \widemwtn{u_g^2}
							   }$  	\\
 \		& $+ \frac\nlcoef {2a} \mwtn v_g \tan\lat \bracket{
		  \widemwtn {u_g^2} + \widemwtn {v_g^2}           }$ \\
 $\Gamma_{A,g}$ & $\frac\nlcoef {2a\cos\lat} \frac {\Fr^2}{N^2}\left[
		   \widemwtn{u_g \Dz\Psi} \frac{\D^2\mwtn\Psi} {\D\lon\D z}
	          + \widemwtn{v_g \Dz\Psi} \frac{\D^2\mwtn\Psi} 
					{\D\lat\D z} \cos\lat \right.$ \\
 \		& \qquad\qquad $\left.
		  - \Dz {\mwtn\Psi} \Dlon {\widemwtn{u_g\Dz\Psi}}
		  - \Dz {\mwtn\Psi} 
		    \Dlat {\parenth{\widemwtn{v_g\Dz\Psi} \cos\lat}}
						 	      \right]$ \\

 \hline
\end{tabular}
\end{center}
\end{table}

\section{Interaction analysis and horizontal treatment}
\label{sect:interact_analysis}

\subsection{Interaction analysis}
\begin{figure*} [t]
   \begin{center}
   \includegraphics[angle=0,width=0.75\textwidth] {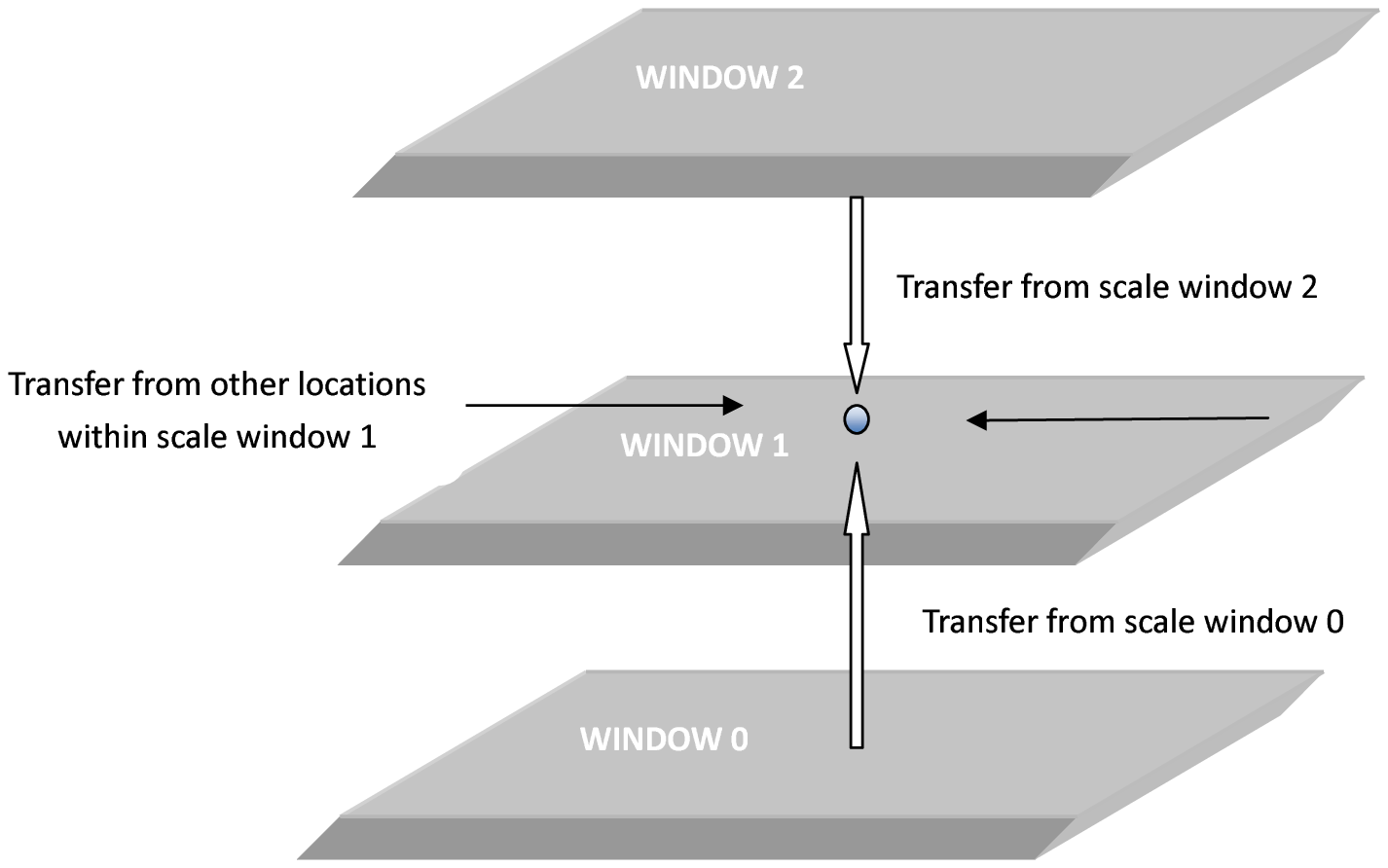}
   \caption
         {A schematic of the canonical energy transfer toward scale window 1.
         \protect{\label{fig:interact}}}
   \end{center}
   \end{figure*}

An energy transfer process toward a certain location in a scale window
involves not only the transfer from outside the window, 
but also those from within. This is a fundamental point where it differs
from that based on the classical Fourier transform or Reynolds
decomposition. Take for an example a transfer\footnote
   {In this section, the dependence on $n$ is kept in the notations.} 
$\Gamma_n^1$ at 
location (step) $n$ in window $1$. As schematized in
Fig.~\ref{fig:interact}, it is the totality of the transfers from 
window 0, window 2, and those from the other different locations (the
sampling space) within the same window. We need to distinguish these
sub-processes in order for the window-window interactions to stand out.

% 
% fig:interact to be inserted here
%

As shown above, all the transfers can be written as a linear combination of
terms in the form
	\begin{eqnarray*}
	\Gamma_n^\win = \mwt \recept n \win \widemwt {pq} n \win.
	\end{eqnarray*}
It therefore suffices to analyze this single term. To make the
presentation easier, we here just pick the particular case $\Gamma_n^1$.
For a detailed treatment, see LR05, section~9. 
Now what we are considering is the transfer
	\begin{eqnarray*}
	\Gamma_n^1 
	&=& \mwt \recept n 1 \widemwt {pq} n 1 = 
	  \mwt\recept n 1  
	  \widemwt {\sum_{\w_1=0}^2 \mwtr p {\w_1}  
		    \sum_{\w_2=0}^2 \mwtr q {\w_2}} n 1	 \\
	&=&
	  \mwt \recept n 1 \bracket{
	      \widemwt {\mwtr p 0 \mwtr q 0} n 1
	    + \widemwt {\mwtr p 0 \mwtr q 1} n 1
	    + \widemwt {\mwtr p 1 \mwtr q 0} n 1
				   }			\\
	&& + \mwt \recept n 1 \bracket{
	      \widemwt {\mwtr p 1 \mwtr q 2} n 1
	    + \widemwt {\mwtr p 2 \mwtr q 1} n 1
	    + \widemwt {\mwtr p 2 \mwtr q 2} n 1	
				  }			\\
	&& + \mwt \recept n 1 \bracket{
	      \widemwt {\mwtr p 0 \mwtr q 2} n 1
	    + \widemwt {\mwtr p 2 \mwtr q 0} n 1
				   }			\\
	&& + \mwt \recept n 1 
	      \widemwt {\mwtr p 1 \mwtr q 1} n 1.
	\end{eqnarray*}
The first two terms represent the energy transfers to scale
window 1 from windows~0 and 2, respectively; write them as 
	    $\Gamma_n^{0\to1}$ and $\Gamma_n^{2\to1}$.
The two scale windows may also combine to contribute to $\Gamma_n^1$,
though generally the contribution is negligible;
this makes the third term, or $\Gamma_n^{0\oplus2\to1}$ for short.
The last term, 
$\Gamma_n^{1\to1} = \mwt\recept n 1 \widemwt {\mwtr p 1 \mwtr q 1} n 1$ is the
transfer from window 1 itself. 
The major purpose of interaction analysis is, for scale window 1,
to select $\Gamma_n^{0\to1}$ and $\Gamma_n^{2\to1}$ out of 
$\Gamma_n^1$. 

For canonical transfers to other scale
windows, the analysis results are referred to Table~\ref{tab:interaction}.

\begin{table}
 \begin{center}
 \caption{Interaction analysis for $\Gamma^0$, $\Gamma^1$, and $\Gamma^2$.
		\label{tab:interaction}}
\begin{tabular}{c c c c c}
 \hline
 \hline	
 $\Gamma^2$	& $\Gamma^{0\to2}$  & $\Gamma^{1\to2}$  &
		  $\Gamma^{0\oplus1\to2}$  & $\Gamma^{2\to2}$	\\
 $\Gamma^1$	& $\Gamma^{0\to1}$  & $\Gamma^{2\to1}$  &
		  $\Gamma^{0\oplus2\to1}$  & $\Gamma^{1\to1}$	\\
 $\Gamma^0$	& $\Gamma^{1\to0}$  & $\Gamma^{2\to0}$  &
		  $\Gamma^{1\oplus2\to0}$  & $\Gamma^{0\to0}$	\\
 Remark:        & instability related & instability related 
		  & usually negligible &\ 	\\
 \hline
\end{tabular}
\end{center}
\end{table}

% \newcommand{\mwt}[3] {{\widehat {#1}_{#2}^{\sim{#3}}}}
% \newcommand{\widemwt}[3] {{\widehat { \left(#1\right)  }_{#2}^{\sim{#3}}}}
%
% \newcommand{\mwtr}[2] {{{#1}^{\sim #2}}}
%	% multiscale window reconstruction

\subsection{Phase oscillation}

The localized multiscale energetics as introduced above may reveal
some spurious high-wavenumber oscillation that must be removed.
This is a fundamental problem with real-valued localized transforms, 
which has been carefully examined 
by \cite{Iima}
		% Iima and Toh (1995) 
in the context of shock waves and wavelet 
analysis. Since this is a technical issue that may prevent one from
making the right interpretation, we here give it a brief introduction; 
details are referred to LR05.

As others, the MWT transform coefficients contain phase information, and 
so do the resulting multiscale energies, which are essentially the square
of the coefficients. The phase information may not be obvious in the
sampling space of the transform coefficients (with elements labeled by $n$)
because of its discrete nature. But the disguised information may appear in
the horizontal through a mechanism like Galilean transformation. (In the
vertical direction it is negligible because the vertical velocity is
generally very weak for geofluid flows.)
To illustrate, look at (\ref{eq:mwt}) that defines the MWT.
The characteristic frequency is $f_c \sim 2^{j_2}$ cycles over the 
time duration. Let the time step size be $\Delta t$, then 
$f_c \sim 1/\Delta t$.
%
% (Recall the signals are equally sampled on $2^{j_2}$ points in time.) 
%
For a flow with speed $u_0$, the oscillation in time with $f_c$
will result in a oscillation in the horizontal with a wavelength 
$\sim u_0 \Delta t$, i.e., a wavenumber $k_c \sim \frac 1 {u_0\Delta t}$. 
Let the mesh size be $\Delta x$. 
For a model to be numerically stable, the CFL condition requires that
$\Delta t < \Delta x / u_0$. So the spurious oscillation has a 
wavenumber $k_c \sim O(\frac 1 {\Delta x})$.

The phase oscillation is a problem rooted in the nature of localized 
transforms. In our case, fortunately, it is always around the highest 
wavenumbers or smallest spatial scales in the spectrum, 
and is hence very easy to be removed using, for example, a 2D large-scale
window reconstruction (like a horizontal low-pass filtering).
	% without incurring those effect such as aliasing. 
This is in contrast to wavelet analysis:
the larger the scale for a transform coefficient, 
the larger the scale for the spurious oscillation (see \cite{Iima}).

\trackchanges{
In real applications the spurious oscillation may not show up, just as in
the MJO case which we will demonstrate in the following section.
But in some unusual cases this could cause severe errors. 
We have shown such an example before in LR05 (see the Fig.~2 therein).
The analysis is with a simulation of an observed meandering 
in the Iceland-Faeroe frontal region in August 1993. The mesh grid 
has a spacing $\Delta x = \Delta y = 2.5\ {\rm km}$, and the time stepsize
is 1800~s. The time series for the multiscale energetics analysis has
a sampling interval of $10\Delta t$. So, by the above argument, the phase
oscillation, if existing, will have a wavelength less than $10\times \Delta
x = 25~km$. 
Indeed, as shown in the Fig.~2a of LR05,
the computed canonical transfer of APE is buried in oscillatory 
errors, with a wavelength of about 8 grid points or 20~km.
These errors are efficiently removed 
through a 2D multiscale window reconstruction with a scale of 25~km;
the resulting transfer is shown in their Fig.2b.
(This can also be achieved efficiently using the 
traditional 2D low-pass filters.)
}

\section{Exemplification with the Madden-Julian Oscillation}
\label{sect:example}

The above formalism has been validated in previous publications and
has seen its success in different real applications. This section is a
demonstration of how it may be applied, with the Madden-Julian Oscillation 
(MJO) as an example. Note here it is not our intention to 
\trackchanges{perform a comprehensive analysis of the MJO energetics},
which will be carefully explored in a forthcoming study.

\begin{figure*}	[h]
	\begin{center}
	\includegraphics[angle=0, width=0.35\textwidth] {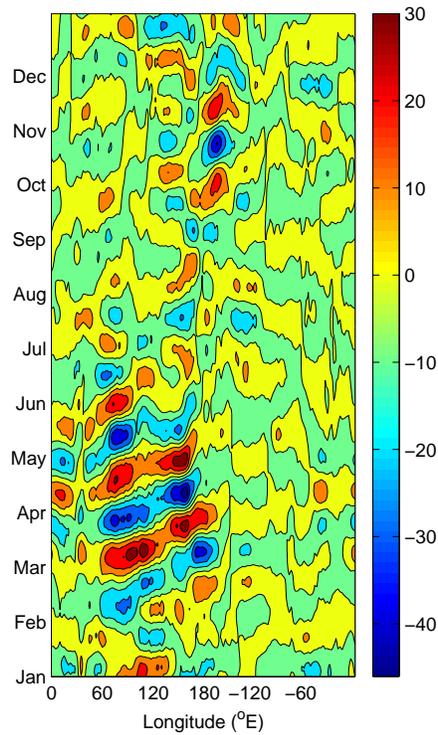}
	\caption{
	The 32-64 day scale window reconstruction for the 1997 OLR anomaly 
	in the tropical area (averaged over 10$^o$S-10$^o$N). 
	Units: $W m^{-2}$.
	\protect{\label{fig:MJO}}}
	\end{center}
	\end{figure*}

MJO is a coupled convection-circulation phenomenon, manifesting 
itself as a localized structure of enhanced and suppressed 
precipitation propagating in the zonal direction at a speed of 4-8~m/s
(cf.~Fig.~\ref{fig:MJO}). 
% It dominates the intraseasonal variabilities 	% (30-90day)
It is the largest element of intraseasonal variability 
in the tropical atmosphere (\cite{MJO}).
	% (Madden and Julian, 1971).
	% with a horizontal scale up to 12,000-20,000~km.
	% also see Madden and Julian 2005 and the references therein). 
Though extending through the whole tropics, the anomalous rainfall % convection
occurs mainly over the Indian Ocean and Western Pacific Ocean. 
The oscillation has a broadband spectrum between the 30-day and 60-day periods.
It is usually strong in winter and spring and weak in summer.
  %
  % In the vertical, it is enhanced in the upper troposphere, and
  % the wind and temperature fields in the upper and lower troposphere 
  % are out of phase.	% while $w$ is in phase.
  %
By observation it originates over the Western Indian Ocean, becomes
strengthened as it enters the Western Pacific, and dies out east
of the dateline. According to 
	 \cite{Wheeler},
	% Wheeler and Hendon (2004), 
a complete MJO cycle comprises of
8 phases, each corresponding to the position of the center of the anomalous 
rainfall, from Western Indian Ocean to Eastern Pacific Ocean. 
As an intraseasonal phenomenon, MJO bridges the large-scale and 
small-scale motions in the atmospheric spectrum, making an important
component of the atmospheric circulation. 
Various studies have established its connections to 
tropical cyclogenesis, El Ni\~no-Southern Oscillation, 
South Asia monsoon, to name a few (see 
	 \cite{MJ05},
	% Madden and Julian, 2005, 
and the references therein). 

% For exammple, it has been observed that the
% first wet phase corresponds well to the summer monsoon in the South 
% China Sea (Wang and Xie, 1997).

	% * dry/wet phase

With large-scale atmospheric circulation and tropical deep convection 
intricately coupled, MJO provides an excellent example for the study of 
multiscale interaction. 
Analytical investigations of the interaction has been made available
in the systematic work of Majda et al. 
       (e.g., \cite{Majda04}, \cite{Majda16}).
     % (e.g., Majda and Biello 2004; Majda, 2007; Majda and Yang, 2016).
Notice the localized and progressive pattern: 
It makes MJO an ideal testbed for our formalism of multiscale energetics. 
We are therefore using it for our purpose of demonstration.

% MJO: (from wiki)
% * "largest element of the intraseasonal (30-90 day) variability in the
% tropical atmosphere"
% * "discovered in 1971 by Roland Madden and Paul Julian of the American
%  National Center for Atmospheric Research (NCAR)"
%* "a large-scale coupling between atmospheric circulation and tropical deep
%   convection"
% * "unlike the standing pattern like ENSO, the MJO is a traveling pattern
%  that propagates eastward at 4-8 m/s, through the atmos avove the warm
%  parts of the Indian and Pacific oceans"
%* "manifests itself most clearly as anonomous rainfall"
%* "characterized by an eastward progression of large regions of both
%  enhanced and suppressed tropical rainfall"
%
% Local effects:
% * monsoon: "A break in the Asian monsoon, normally during the month of
%  July, has been attributed to MJO, after its enhanced phase moves off to
%  the east of the region into the tropical Pacific ocean."
% * tropical cyclogenesis
%
%   (typhoon genesis: monsoon valley(?) in Pacific?  easterly in Atlantic)

The data we are using include those from the ECMWF Re-Analysis Interim
(ERA-Interim)\footnote
   {http://www.ecmwf.int/en/research/climate-reanalysis/era-interim}
daily products (wind, temperature, and geopotential height), 
and the series of the real-time multivariate MJO (RMM) 
	(\cite{Wheeler}).
	% (Wheeler and Hendon, 2004). 
They have a spatial resolution of $2.5^o \times 2.5^o$ and
span from 1988 through 2010. 
The vertical temperature profile, $\bar T = \bar T(p)$, which is needed in the
application, is obtained by taking the time mean of $T$, 
followed by an averaging over all the $p$-planes.

To begin, we need to demarcate the scale windows.
The problem forms a natural three-window decomposition: large-scale
variabilities, MJO, and synoptic processes. 
We choose an MJO window of 32-64 days, since in the analysis a power of 2
for a window bound is required.

% We choose 
%		$j_0=8$,
%		$j_1=9$,
%		$j_2=13$,
% which makes an MJO window of 32-64 days. 
% (We have used the fact that the function space formed with 
% a scaling basis with level $j$ contains features
% with scale level up to $j$ but not including $j$.)

% correspond respectively to processes of longer than 128 days, 
% 32-64 days, and 1-16 days, respectively. The second scale window 
% is intended for MJO.  It may be more appropriate to choose
% 32-64 days (a power of 2 is required), but just in case there are some
% events outside that range, we pick 16-128 days. Anyhow, the variabilities
% within the intraseasonal window need not be MJO ---MJO will be identified 
% from that window with the RMM index. 

\begin{figure*}	[h]
	\begin{center}
	\includegraphics[angle=0, width=0.75\textwidth] {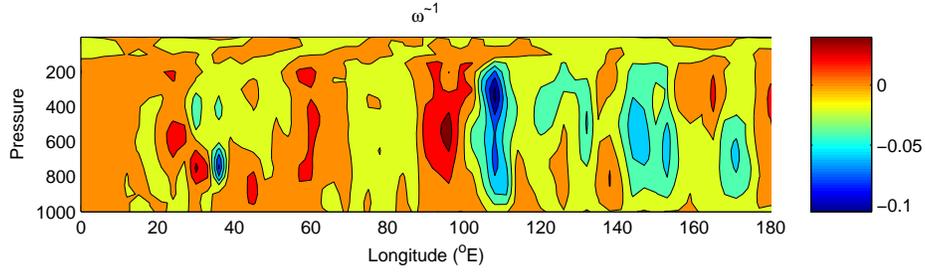}
	\caption{
	The equatorial $\w=dp/dt$ on December 16, 1996, reconstructed on 
	32-64 day scale window. Note the up-westward tilting pattern east of
	the maritime continent.
	\protect{\label{fig:MJO_reconst}}}
	\end{center}
	\end{figure*}

We choose a strong MJO event on December 16, 1996, for our exemplification 
purpose. The RMM index is 2.05, corresponding to phase 5 
(where the convection center is over the maritime continent). 
Using the above parameters, a straightforward application to the 
outgoing longwave radiation (OLR) in the tropical region (averaged 
between 10$^o$S and 10$^o$N) immediately yields an 
MJO window reconstruction (Fig.~\ref{fig:MJO}). From it the 
eastward propagation and its seasonal variation are clearly seen.
Likewise, velocity and temperature can be reconstructed. Particularly,
$\mwtr u 1$, $\mwtr \w 1$, and $\mwtr T 1$ have on the zonal cross section
an up-westward tilting pattern, as identified earlier on
(e.g., \cite{Moncrieff}); see Fig.~\ref{fig:MJO_reconst}. 

\begin{figure*}	[h]
	\begin{center}
	\includegraphics[width=0.75\textwidth] {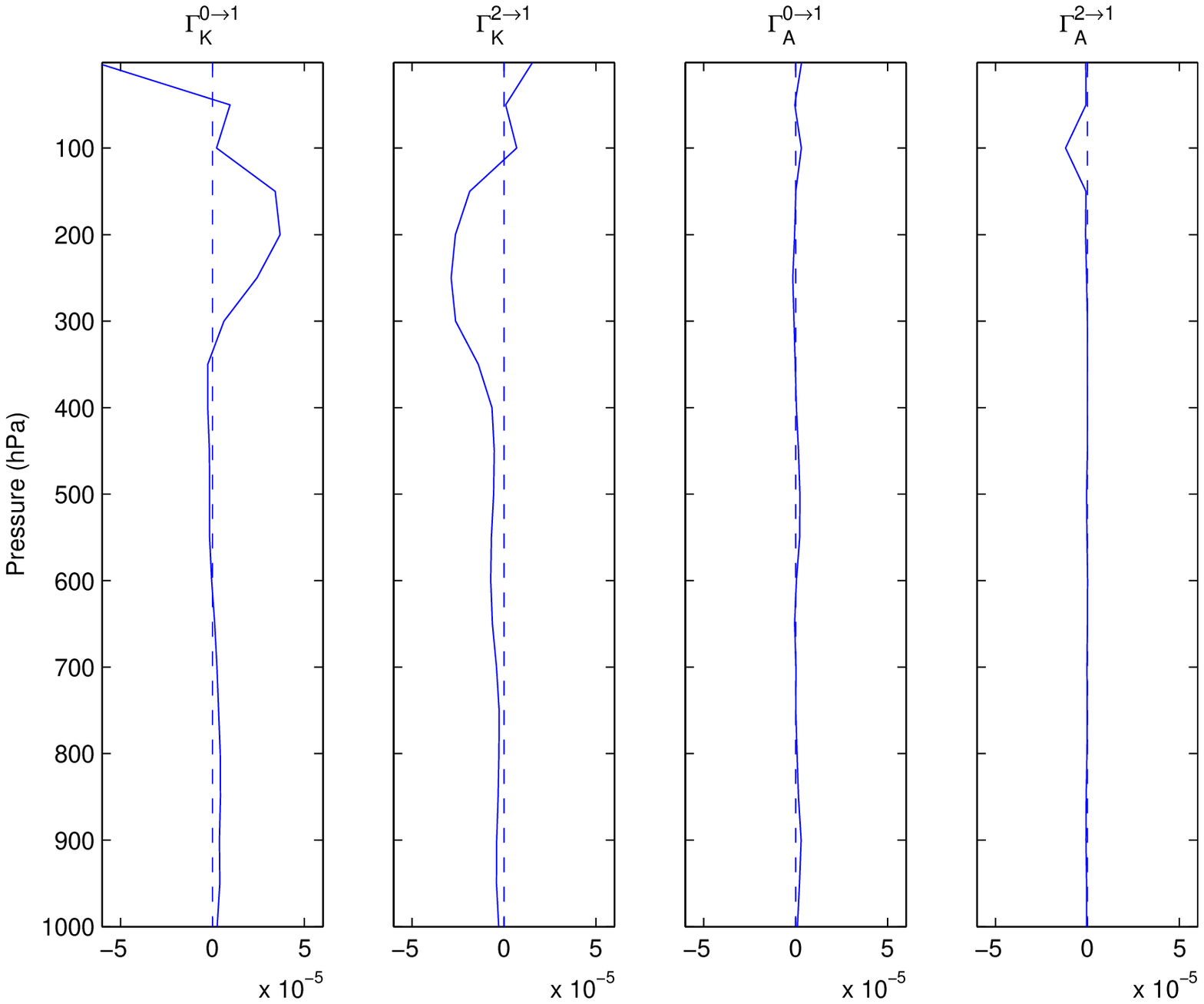}
	\caption{
	$\Gamma_K^{0\to1}$, and $\Gamma_K^{2\to1}$,
	$\Gamma_A^{0\to1}$, and $\Gamma_A^{2\to1}$ averaged between
	10$^o$S-10$^o$N and 0$^o$E-$180^o$E. Units: $m^2/s^3$.
	\protect{\label{fig:MJO_transfer}}}
	\end{center}
	\end{figure*}

Show in Fig.~\ref{fig:MJO_transfer} are the vertical distributions of the
canonical transfers to the MJO window averaged over the tropical region 
(10$^o$S-10$^o$N) between 0$^o$E-180$^o$E. From the kinetic transfers,
$\Gamma_K^{0\to1}$ is on the whole positive, while $\Gamma_K^{2\to1}$ is
negative. That is to say, $\Gamma_K$ is downscale. In contrast, its
potential energy counterpart tends to be more irregularly distributed, and,
besides, is one order smaller.
Though this is just for one particular day only, the long time mean also 
has the trend. This is in opposite to that for the the mid-latitude 
paradigm, where the canonical APE transfer is downscale while the canonical
KE transfer is upscale (\cite{Saltzman70}). 
From the figure the transfer center is located in the upper troposphere
around 200 hPa, in agreement with the previous studies (e.g, \cite{Hsu}).

\begin{figure*}	[h]
	\begin{center}
	\includegraphics[width=0.75\textwidth] {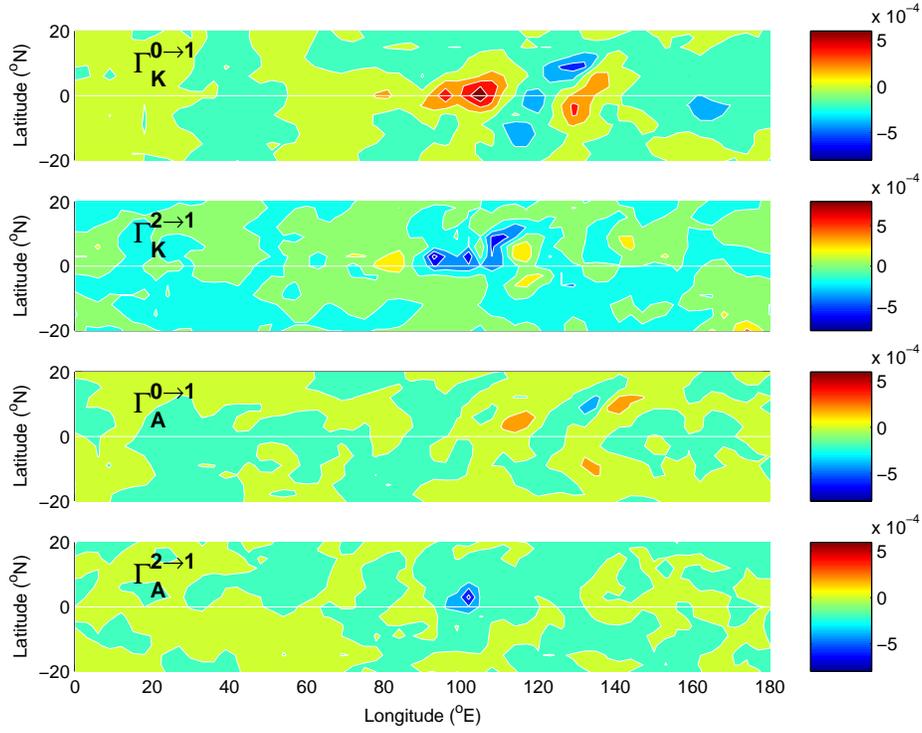}
	\caption{
	$\Gamma_K^{0\to1}$, $\Gamma_K^{2\to1}$,
	$\Gamma_A^{0\to1}$, and $\Gamma_A^{2\to1}$ at
	100~hPa (units: $m^2/s^3$).
	\protect{\label{fig:MJO_transfer_hori}}}
	\end{center}
	\end{figure*}

To examine the the horizontal distributions of instability centers, in 
Fig.~\ref{fig:MJO_transfer_hori} we draw the maps of the 
canonical transfers at 200~hPa.
We see that they are mainly distributed between 
100-140$^o$E, i.e., the maritime continent. This is, of course, 
in agreement with the phase where MJO lies at that time.

We emphasize again that it is not our intention to study the MJO dynamics
here. We just pick for the demonstration purpose such an example at such an
instance. It is seen that, through a straightforward application, one
immediately obtains a bunch of maps of the multiscale energetics that 
reflect the underlying internal dynamics, and these energetics agree well
with the previous studies. A detailed study of MJO the intraseasonal mode
requires a statistical examination of the resulting energetics; we will see
that later in Lu et al.\footnote
  {Lu, H.C., Y.B. Zhao, and X.S. Liang: 
   The multiscale MJO energetics (in preparation).
  }.

\section{Conclusions and discussion}	\label{sect:summary}

\trackchanges{
Multiscale energetics diagnostics are important in that they 
provide an approach to the fundamental problems of atmospheres and
oceans like mean flow-disturbances interaction, instability and
disturbance growth, etc., as identified in the National Report of 
Lindzen and Farrell\cite{Lindzen}.
  %
  % 1983-1986 National Report to International Union of Geodesy and 
  % Geophysics (IUGG).
  %
  % \cite{Lindzen} identified 6 topics of most inerest, among which 
  % the following two
  %	\begin{itemize}
  %	\item Diagnostics for the interaction among disturbances and the
  %          mean flow
  %	\item Non-traditional approaches to instability and disturbance
  %	      growth
  %	\end{itemize}
  % are directly related to this study.
  % hence the fundamental dynamics, within the geophysical fluid flows, 
  %
Their importance is also seen in the potential role that they may play in 
the major engineering problems such as eddy transport parameterization 
(e.g., 
	\cite{Gent},
	\cite{Greatbatch},
	\cite{Visbeck},
	\cite{DMarshall10}),
	% Gent and McWilliams 1990; 
	% Greatbatch 1998; 
	% Visbeck et al., 1997; 
	% Marshall and Adcroft 2010).
turbulence and feedback closure (e.g., \cite{Jin}), etc. }
Based on the new analysis machinery namely multiscale window transform
(MWT), which is capable of orthogonally decomposing a function space into a 
direct sum of several subspaces while retaining the local information in
the resulting transform coefficients,
we have given a comprehensive derivation of the multiscale energetics
for the atmosphere, with respect to both the primitive equation
and model quasi-geostrophic model. By taking advantage of the nice properties of 
the MWT, an ``atomic'' reconstruction of the fluxes on the 
multiscale windows allows for a unique separation of the inter-scale
transfer from the nonlinearly intertwined energetics. The resulting
transfer bears a Lie bracket form, reminiscent of the Poisson bracket in
the Hamiltonian dynamics; for this reason, we call it {\it canonical
transfer}. A canonical transfer sums to zero over scale windows,
indicating that it is a mere redistribution of energy 
among the scale windows, without generating or destroying energy as a whole. 

The multiscale atmospheric kinetic energy (KE) 
and available potential energy (APE) equations are thence derived. 
By classification, a multiscale energetic cycle comprises of the following
processes: KE transport, APE transport, pressure work, buoyancy conversion, 
work done by external forcing and diabatic and frictional processes 
in the respective scale windows, and the interscale canonical transfers
of KE and APE, which have been shown to correspond to the
barotropic and baroclinic instabilities\cite{LR07}.
Note that a buoyancy conversion takes place in an individual window only,
bridging the two types of energy namely KE and APE. It does not involve the
process among different scale windows, and hence basically is not related
to instabilities, although traditionally it has been used to diagnose
baroclinic instabilities. A brief application of the formalism is 
exemplified with the Madden-Julian Oscillation.

Also derived are the multiscale KE and APE equations 
for quasi-geostrophic flows and, for completeness, 
those for oceanic circulations.
It should be cautioned that,
since what we talk about are four-dimensional energy distribution 
and evolution, the term ``energy'' in this study is, in a strict
sense, ``energy density.'' The abuse of terminology will not cause
confusion as it is clear in the context.

\trackchanges{
It should be mentioned that the definition of APE is still an active
arena of research; a recent review can be found in \cite{Tailleux}. 
In the present formalism, APE is defined as that in
\cite{Lorenz}, which takes a quadratic form.
However, it has been argued that it is generally not quadratic, 
if the 1D reference hydrostaic thermodynamic profile is achieved by 
adiabatic rearrangement of the existing 3D state 
(e.g., \cite{Holliday}, \cite{Winters95}, \cite{Winters13}).
This raises an issue about how to handle an APE in non-quadratic form 
in the multiscale formalism. Recall that, in this study, 
central at the multiscale energy representation is the Parseval 
relation, while the relation works only for quadratic properties.
For a non-quadratic APE, the problem may need to be considered
from a more fundamental point of view.
We will leave that to future discussions.

}

Notice that presented in this study is about the energetics based on 
a three-scale window decomposition. It is straightforward to extend the
formalism to four, five, or more scale windows; the resulting energy equations
are the same in form. One may equally reduce the number of windows to two.

We remark that there is a well-known apparatus in achieving a two-scale
decomposition in atmospheric research, that is, the decomposition
through taking transformed Eulerian mean (\cite{Andrews76},
\cite{McIntyre};
	% Andrews and McIntyre, 1976; 
	% McIntyre, 1980; 
also see 
	\cite{Plumb05}, \cite{Buhler}).
	% Plumb and Ferrari, 2005; B\"uhler, 2009). 
Formalisms of two-scale energetics have been established with 
the theory (e.g., \cite{Plumb83}). But how these formalisms may be related to
that in this study is yet to be carefully examined.

% "Theoretical formulations of eddy-mean-flow interaction under
% near-adiabatic conditions are most elegant, and most powerful, within an
% isentropic-or isopynal coordinate framework (e.g., Andrews 1983; 
% Greatbatch 1998)."

In LR05, there is also a brief touch on multiscale enstrophy analysis, 
which makes the whole a ``localized multiscale energy and 
vorticity analysis,'' or MS-EVA for short. 
Since the multiscale enstrophy equation is closely related to 
an important concept in dynamic meteorology, namely, the Eliassen-Palm flux
	 (\cite{Eliassen}, \cite{Buhler}, \cite{Vallis}), 
	% (Eliassen and Palm, 1961; B\"uhler 2009; Vallis 2006), 
\trackchanges{which has been extensively employed in wave-activity diagnosis}
and certainly deserves a detailed study for its own sake 
	 (e.g., \cite{JMarshall84}, \cite{Plumb86}, \cite{Rhines},
	\cite{Nakamura}, \cite{Takaya}), 
	% (e.g., Marshall, 1984; Plumb, 1986; Rhines and Holland, 1979), 
we will defer it to another investigation in the near future.

%%%%%%%%%%%%%%%%%%%%%%%%%%%%%%%%%%%%%%%%%%%%%%%%%%%%%%%%%%%%%%%%%%%%%
% ACKNOWLEDGMENTS
%%%%%%%%%%%%%%%%%%%%%%%%%%%%%%%%%%%%%%%%%%%%%%%%%%%%%%%%%%%%%%%%%%%%%
%
\acknowledgments
Thanks are due to ECMWF for providing the ERA-Interim data.
This research was supported 
by the National Science Foundation of China under Grant No. 41276032, 
by Jiangsu Provincial Government through the 2015 Jiangsu Program for 
Innovation Research and Entrepreneurship Groups and through the Jiangsu
Chair Professorship, and by the State Oceanic Administration 
through the National Program on
Global Change and Air-Sea Interaction (GASI-IPOVAI-06).

%%%%%%%%%%%%%%%%%%%%%%%%%%%%%%%%%%%%%%%%%%%%%%%%%%%%%%%%%%%%%%%%%%%%%
% APPENDIXES
%%%%%%%%%%%%%%%%%%%%%%%%%%%%%%%%%%%%%%%%%%%%%%%%%%%%%%%%%%%%%%%%%%%%%
%
% Use \appendix if there is only one appendix.
%\appendix

%%%%%%%%%%%%%%%%%%%%%%%%%%%%%%%%%%%%%%%%%%%%%%%%%%%%%%%%%%%%%%%%%%%%%%%%%%%%%%
\appendix
\section{A glossary of notations}
% \appendixtitle{A glossary of notations}	
%%%%%%%%%%%%%%%%%%%%%%%%%%%%%%%%%%%%%%%%%%%%%%%%%%%%%%%%%%%%%%%%%%%%%%%%%%%%%%

\begin{itemize}
\item $\grad$: 3D gradient operator, 
	$\grad = \ve e_\lon\Dlon\ + \ve e_\lat\Dlat\  + \ve e_p\Dp\ $ 
\item $\gradh$: horizontal gradient operator (horizontal component of $\grad$)
\item $\ve v$: velocity: for atmospheres, $\ve v = (u,v,\w)$, $\w=dp/dt$;
			for oceans,  $\ve v = (u,v,w)$.
\item $\vh = (u,v)$ 
\item $\phi$: scaling function
\item $\mwtn \pi$: MWT of some property $\pi$ at step $n$ on window $\win$;
	dependence on $n$ is suppressed when no confusion arises
\item $\pi^{\sim\win}$: window $\win$-filtered $\pi$ 
	(multiscale window reconstruction of $\pi$ on window $\win$)
\item $\pi_n^\win$: notation of some property at step $n$ on window $\win$;
	$n$ is suppressed when no confusion arises.
\item $\bar T = \bar T(p)$:  mean temperature profile (averaged over the
	$p$-plane and time)
\item $T$: departure from $\bar T$
\item $\alpha$: specific volume
\item $\gp$: geopotential function
\item $Z$: geopotential height $(Z=\Phi / g)$
\item $R$: specific gas constant (in $\rm J\cdot kg^{-1} K^{-1}$)
\item $c_p = 1.005\times10^3$: specific heat capacity of air (in J/kg.K)
		for isobaric processes
\item $c_v$: specific heat capacity of air for isochoric processes
\item $f$: Coriolis parameter
\item $\beta$: meridional gradient of $f$
\item $L$: Lapse rate ($L = -\Dz {\bar T}$)
\item $L_d$: Lapse rate of dry air ($L_d = g/c_p$)
\item $a$: radius of Earth
\item $(\lon, \lat, r)$: the spherical coordinates
\item $p$: pressure coordinate
\item $z = r-a$; $dx=a\cos\lat d\lon$, $dy=ad\lat$
\item $(\tilde x, \tilde y, \tilde z)$: Cartesian coordinates
\item $\ve i, \ve j, \ve k$: unit vectors for the cartesian coordinate system
\item $\ve e_\lon, \ve e_\lat, \ve e_z$: unit vectors for spherical
		coordinate system
\item $\ve e_\lon, \ve e_\lat, \ve e_p$: unit vectors for the isobaric spherical
		coordinate system
\item $g$: acceleration due to gravity
\item $(h_\lon, h_\lat, h_z)$: Lam\'e's coefficients
\item $\bar\rho = \bar\rho(z)$:  stationary density profile (ocean)
\item $\rho$: density perturbation with $\bar\rho$ removed (ocean)
\item $\rho_0$: chosen to be 1025 (kg/m$^3$) here (ocean)
\item $N = N(z) = \sqrt{-\frac g {\rho_0} \Dz {\bar\rho}}$: 
	buoyancy frequency (ocean)
\item $P$: dynamic pressure, i.e., pressure with 
	$\bar P(z) = \bar P_0 - \int_0^z \bar\rho g dz$ removed (ocean)
\item $c = \frac g {\bar T(g/c_p - L)}$ (atmosphere);
      $c = \frac {g^2} {\rho_0^2 N^2}$ (ocean)
\item $\ve Q$: flux
\item $\Gamma$: canonical transfer
\item $A$: available potential energy
\item $K$: kinetic energy
\item $b$: buoyancy conversion rate
\item $F_h$: friction/external forcing in horizontal direction
\item $F_z$: friction/external forcing in vertical direction
\item $F_p$: friction/external forcing in $p$ direction
\item $\psi$: streamfunction
\item $\ve v_g$: geostrophic velocity ($=\ve k \times \gradh\psi$)
\item $\Fr$: rotational internal Froude number
\item $\eps$: Rossby number $(=\frac 1 {f_0t_0})$
\item $\nlcoef = \frac {U_0t_0} {L_0}$
\item ${\mathscr L} = \Dt\ + \ve v_g \cdot \gradh = \Dt\ + J(\psi,\ )$
\end{itemize}

%%%%%%%%%%%%%%%%%%%%%%%%%%%%%%%%%%%%%%%%%%%%%%%%%%%%%%%%%%%%%%%%%%%%%%%%%%%%%%
% \appendix[B]	
% \appendixtitle{Expansion of $\grad\cdot(\ve v\ve v)$ in spherical coordinates}
\section{Expansion of $\grad\cdot(\ve v\ve v)$ in spherical coordinates}
%%%%%%%%%%%%%%%%%%%%%%%%%%%%%%%%%%%%%%%%%%%%%%%%%%%%%%%%%%%%%%%%%%%%%%%%%%%%%%

\begin{figure*}	[t]
	\begin{center}
	\includegraphics[angle=0, width=0.65\textwidth]
	{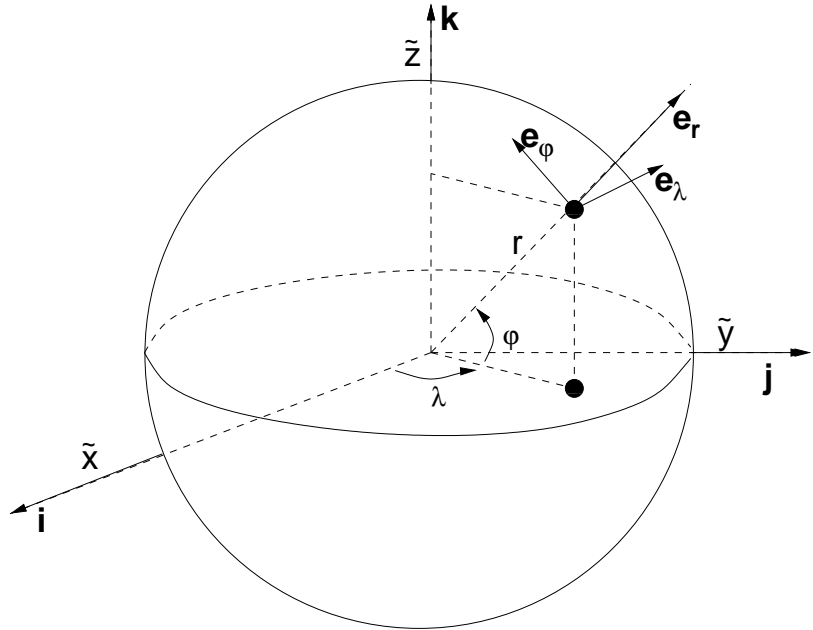}
	% \appendcaption{
	\caption{
	Spherical coordinate frame.
	\protect{\label{fig:spheric_frame}}}
	\end{center}
	\end{figure*}

To compute the canonical transfer (\ref{eq:Gamma1}), we are required to
evaluate explicitly $\grad\cdot (\ve v \vh)$ in the spherical coordinate
system ($\lon$, $\lat$, $r$), which are connected with the cartesian
coordinates $(\tilde x, \tilde y, \tilde z)$ as follows:
	\begin{eqnarray}
	&& \tilde x = r\cos\lat \cos\lon,	\\
	&&\tilde y = r\cos\lat \sin\lon,	\\
	&& \tilde z = r \sin\lat.
	\end{eqnarray}
Here the over-tilde is employed to avoid confusing with $z$ which will be 
reserved for height measuring from the earth surface: 
$z = r - a$, with $a$ being the radius of Earth. 
Besides, in meteorology, $dx$ and $dy)$ are usually reserved for 
$a\cos\lat d\lon$ and $a d\lat$, respectively.
From the position vector
$\ve x = \tilde x \ve i + \tilde y \ve j + \tilde z \ve k$
it is easy to find the Lam\'e's coefficients as follows
(cf.~Fig.~\ref{fig:spheric_frame}):
	\begin{eqnarray}
	&& h_\lon = \abs{\DD {\ve x} {\lon}} = r\cos\lat, \\
	&& h_\lat = \abs{\DD{\ve x} {\lat}} = r, \\
	&& h_z = \abs{\DD{\ve x} {z}} = 1.
	\end{eqnarray}

%
%	Figure fig:spheric_frame here
%

With the shallow water approximation, $r\approx a = \const$. So
	\begin{eqnarray}
	&&\grad\cdot (\ve v\ve v)
	= \grad\cdot [\ve v (u\ve e_\lon  + v\ve e_\lat + w\ve e_z )] \cr
   	&&= 
	  \frac 1 {a\cos\lat} 
	  \Dlon {[u(u\ve e_\lon + v\ve e_\lat + w\ve e_z)]}
	+ \frac 1 {a\cos\lat}
	  \Dlat {[v (u\ve e_\lon + v\ve e_\lat + w\ve e_z) \cos\lat]}	\cr
	&&\quad
	+ \Dz {[w (u\ve e_\lon + v\ve e_\lat + w\ve e_z)]}.
					\label{eq:gradvv}
	\end{eqnarray}
And, particularly,
	\begin{eqnarray}
	&&\grad\cdot (\ve v \vh)
	= \grad\cdot [\ve v (u\ve e_\lon  + v\ve e_\lat)] \cr
   	&&= 
	  \frac 1 {a\cos\lat} 
	  \Dlon {[u(u\ve e_\lon + v\ve e_\lat)]}
	+ \frac 1 {a\cos\lat}
	  \Dlat {[v (u\ve e_\lon + v\ve e_\lat) \cos\lat]}	\cr
	&&\quad
	+ \Dz {[w (u\ve e_\lon + v\ve e_\lat)]}.
	\end{eqnarray}

We need to evaluate $\Dlon {\ve e_\lon}$, $\Dlat {\ve e_\lon}$, etc.

There are several ways to achieve the evaluation. One may do it by directly
taking the limit 
  $\Dlon{\ve e_\lon} = \lim_{\Delta\lon\to0} 
	\frac {\Delta\ve e_\lon} {\Delta\lon}.$
Another way is to first connect $(\ve e_\lon, \ve e_\lat, \ve e_z)$
with $(\ve i, \ve j, \ve k)$, then take the derivatives.
Besides, one may take advantage of the properties such as:
	\begin{eqnarray*}
	\ve e_\lon \cdot \ve e_\lat = 1 
	\Longrightarrow \D \ve e_\lon \cdot \ve e_\lon = 0
	\Longrightarrow \D\ve e_\lon \perp \ve e_\lon.
	\end{eqnarray*}

From Fig.~\ref{fig:spheric_frame}, it is easy to find that
% (e.g., Jacobson, 2005?)
	\begin{eqnarray}
	&& \ve e_\lon = - \sin\lon \ve i + \cos\lon \ve j,	\\
	&& \ve e_\lat = -\sin\lat\cos\lon \ve i - \sin\lat\sin\lon \ve j
		     + \cos\lat \ve k,	\\
	&& \ve e_z = \cos\lat\cos\lon \ve i + \cos\lat\sin\lon \ve j
		     + \sin\lat \ve k.
	\end{eqnarray}
Inverting,
	\begin{eqnarray}
	&&\ve i = -\cos\lon\sin\lat \ve e_\lat + \cos\lon\cos\lat \ve e_z
		- \sin\lon \ve e_\lon,	\\
	&&\ve j = -\sin\lon\sin\lat \ve e_\lat + \sin\lon\cos\lat \ve e_z
		+ \cos\lon \ve e_\lon,	\\
	&&\ve k = \cos\lat \ve e_\lat + \sin\lat \ve e_z.
	\end{eqnarray}
So
	\begin{eqnarray}
	&& \Dlon {\ve e_\lon} = -\cos\lon \ve i - \sin\lon \ve j
			   = \sin\lat \ve e_\lat - \cos\lat \ve e_z, \\
	&& \Dlat {\ve e_\lon} = 0,	\\
	&& \Dz {\ve e_\lon} = 0, 	\\
	&& \Dlon {\ve e_\lat} = \sin\lat\sin\lon \ve i - \sin\lat\cos\lon \ve j
			   = -\sin\lat \ve e_\lon,	\\
	&& \Dlat {\ve e_\lat} = -\cos\lat\cos\lon \ve i 
			     - \cos\lat\sin\lon \ve j
			     - \sin\lat \ve k
			   = - \ve e_z,		\\
	&& \Dz {\ve e_\lat} = 0,			\\
	&& \Dlon{\ve e_z} = -\cos\lat\sin\lon \ve i +\cos\lat\cos\lon \ve j
		       = \cos\lat \ve e_\lon,	\\
	&& \Dlat{\ve e_z} = -\sin\lat\cos\lon \ve i - \sin\lat\sin\lon \ve j
				+ \cos\lat \ve k
			= \ve e_\lat,		\\
	&& \Dz {\ve e_z} = 0.
	\end{eqnarray}
Also one may obtain:
	\begin{eqnarray}
	&& \dt {\ve e_\lon} = \frac u {a\cos\lat} (\ve e_\lat \sin\lat 
				- \ve e_z \cos\lat),	\\
	&& \dt {\ve e_\lat} = \frac {u\tan\lat} a \ve e_\lon 
			 - \frac va \ve e_z,	\\
	&& \dt {\ve e_z} = \frac ua \ve e_\lon + \frac va \ve e_z.
	\end{eqnarray}

With the above results, (\ref{eq:gradvv}) now can be expanded as
	\begin{eqnarray}
	\grad\cdot (\ve v \ve v)
	&=& \frac 1 {a\cos\lat} 
	  \bracket{ \Dlon {u^2} \ve e_\lon + \Dlon {uv} \ve e_\lat
			+ \Dlon {uw} \ve e_z
		    + u^2 \Dlon {\ve e_\lon} + uv \Dlon {\ve e_\lat}
			+ uw \Dlon {\ve e_z}
		  }	\cr
	&+&
	  \frac 1 {a\cos\lat}
	  \left[\Dlat{(vu\cos\lat)} \ve e_\lon +
		   \Dlat{(v^2\cos\lat)} \ve e_\lat +
		   \Dlat {(vw \cos\lat)} \ve e_z  \right.	\cr
	&\ &\qquad
		+  \left.
		   vu\cos\lat \Dlat {\ve e_\lon} +
		   v^2\cos\lat \Dlat {\ve e_\lat} +
		   vw\cos\lat \Dlat {\ve e_z}	
		   \right]	\cr
	&+&
	  \bracket{\Dz{wu} \ve e_\lon +
		   \Dz{wv} \ve e_\lat +
		   \Dz {w^2} \ve e_z +
		   wu \Dz {\ve e_\lon} +
		   wv \Dz {\ve e_\lat} +
		   w^2 \Dz {\ve e_z}	
		   }
	\end{eqnarray}
Or,
%	\begin{linenomath*}	% force a linenumber for the equation
	\begin{eqnarray}
	&& \grad\cdot (\ve v \ve v) 
	=
	\cbrace{
	   \frac1{a\cos\lat}\bracket{
		 \Dlon{u^2} - uv \sin\lat + uw \cos\lat
		+ \Dlat {(vu\cos\lat)}
				    }  + \Dz {wu}
		          } \ve e_\lon		\cr
	&&\quad +
	\cbrace{
	   \frac1{a\cos\lat}\bracket{
		 \Dlon{uv} + u^2 \sin\lat + \Dlat {(v^2\cos\lat)}
			+ vw \cos\lat
				    }  + \Dz {wv}
		          } \ve e_\lat		\cr
	&&\quad +
	\cbrace{
	   \frac1{a\cos\lat}\bracket{
		 \Dlon{uw} - u^2 \cos\lat + \Dlat {(vw\cos\lat)}
			- v^2 \cos\lat
				    }  + \Dz {w^2}
		          } \ve e_z.	\label{eq:gradvv2}
	\end{eqnarray}
%	\end{linenomath*}

One may check that, with the aid of the incompressibility assumption
	\begin{eqnarray*}
	\frac1{a\cos\lat} \Dlon u + \frac1{a\cos\lat} \Dlat {v\cos\lat}
	  + \Dz w = 0,
	\end{eqnarray*}
the above equation is equivalent to
	\begin{eqnarray*}
	&& \parenth{\frac u {a\cos\lat} \Dlon u + \frac va \Dlat u + w\Dz u
		- \frac {uv}a \tan\lat + \frac {uw} a} \ve e_\lon  \cr
	&& + \parenth{\frac u{a\cos\lat}\Dlon v + \frac {u^2}a \tan\lat
		+ \frac va \Dlat v + \frac {vw}a + w\Dz v} \ve e_\lat \cr
	&& + \parenth{\frac u{a\cos\lat}\Dlon w + \frac va \Dlat w + w\Dz w
		- \frac {u^2+v^2} a}\ve e_z,
	\end{eqnarray*}
which is precisely the advection part in the 
non-approximated momentum equations in spherical coordinates. 
Eq.~(\ref{eq:gradvv2}) is thence verified.

As a particular case,
	\begin{eqnarray}
	\grad\cdot (\ve v \vh)
	&=& \frac 1 {a\cos\lat} 
	  \bracket{ \Dlon {u^2} \ve e_\lon + \Dlon {uv} \ve e_\lat
		    + u^2 \Dlon {\ve e_\lon} + uv \Dlon {\ve e_\lat} 
		  }	\cr
	&+&
	  \frac 1 {a\cos\lat}
	  \left[\Dlat{(vu\cos\lat)} \ve e_\lon +
		   \Dlat{(v^2\cos\lat)} \ve e_\lat 
		   \right.	\cr
	&\ &\qquad
		+  \left.
		   vu\cos\lat \Dlat {\ve e_\lon} +
		   v^2\cos\lat \Dlat {\ve e_\lat} 
		   \right]	\cr
	&+&
	  \bracket{\Dz{wu} \ve e_\lon +
		   \Dz{wv} \ve e_\lat +
		   wu \Dz {\ve e_\lon} +
		   wv \Dz {\ve e_\lat}
		   }
	\end{eqnarray}
Or,
	\begin{eqnarray}
	&& \grad\cdot (\ve v \vh) 
	=
	\cbrace{
	   \frac1{a\cos\lat}\bracket{
		 \Dlon{u^2} - uv \sin\lat
		+ \Dlat {(vu\cos\lat)}
				    }  + \Dz {wu}
		          } \ve e_\lon		\cr
	&&\quad +
	\cbrace{
	   \frac1{a\cos\lat}\bracket{
		 \Dlon{uv} + u^2 \sin\lat + \Dlat {(v^2\cos\lat)}
				    }  + \Dz {wv}
		          } \ve e_\lat		\cr
	&&\quad +
	\cbrace{
	   \frac1{a\cos\lat}\bracket{
		 - u^2 \cos\lat 
			- v^2 \cos\lat
				    } 
		          } \ve e_z.	
	\end{eqnarray}
Correspondingly with the incompressibility assumption this is,
	\begin{eqnarray*}
	&& \parenth{\frac u {a\cos\lat} \Dlon u + \frac va \Dlat u + w\Dz u
		- \frac {uv}a \tan\lat} \ve e_\lon  \cr
	&& +\parenth{\frac u{a\cos\lat}\Dlon v + \frac {u^2}a \tan\lat
		+ \frac va \Dlat v + w\Dz v} \ve e_\lat \cr
	&& +\parenth{
		- \frac {u^2+v^2} a}\ve e_z.
	\end{eqnarray*}
%%%%%%%%%%%%%%%%%%%%%%%%%%%%%%%%%%%%%%%%%%%%%%%%%%%%%%%%%%%%%%%%%%%%%%%%%%%%%%

%%%%%%%%%%%%%%%%%%%%%%%%%%%%%%%%%%%%%%%%%%%%%%%%%%%%%%%%%%%%%%%%%%%%%%%%%%%%%%
% \appendix[C]
	% \label{sect:app_qg}
% \appendixtitle{Some quasi-geostrophic results used in the text}
\section{Some quasi-geostrophic results used in the text}
%%%%%%%%%%%%%%%%%%%%%%%%%%%%%%%%%%%%%%%%%%%%%%%%%%%%%%%%%%%%%%%%%%%%%%%%%%%%%%
Using the scaling in section~\ref{sect:QG}, it is easy to have the scaled
inviscid governing equations (\ref{eq:gov1_ocean})-(\ref{eq:gov4_ocean})
as follows
(now all the variables in this appendix are understood as nondimensional):
	\begin{eqnarray}
	&& \eps \Dt\vh + \eps\nlcoef \parenth{\vh \cdot \gradh\vh + w\Dz\vh}
	     + f\ve k \times \vh = - \gradh P	\\
	&& \rho = - \Dz P	\\
	&& \gradh\cdot\vh + \Dz w = 0 \\
	&& \Fr^2\eps\Dt\rho + \Fr^2\eps\nlcoef\parenth{\vh\cdot\gradh\rho 
						  + w\Dz\rho}
	   = N^2w
	\end{eqnarray}
where $f = 1 + \eps\beta y$.

Expanding $P$, $w$, $\vh$, and $\rho$ in the power of $\eps$, as that in 
(\ref{eq:qg_epsP})-(\ref{eq:qg_epsrho}), 
	%\begin{eqnarray*}
	% && P = [P]_0 + \eps [P]_1 + \eps^2 [P]_2...	\\
	% && w = [w]_0 + \eps [w]_1 + \eps^2 [w]_2...	\\
	% && \vh = [\vh]_0 + \eps [\vh]_1 + \eps^2 [\vh]_2...\\
	% && \rho = [\rho]_0 + \eps [\rho]_1 + \eps^2 [\rho]_2...,
	%\end{eqnarray*}
it is easy to show that
	\begin{eqnarray}
	&& [\vh]_0 = \ve k \times \grad [P]_0, 	\label{app:qg1} \\
	&& [w]_0 = 0.				\label{app:qg2} \\
	&& [\rho]_0 = -\Dz {[P]_0}		\label{app:qg3}
	\end{eqnarray}
%
%	\begin{eqnarray}
%	&& \frac{\Fr^2}{N^2}\parenth{\Dt {[\rho]_0} + 
%		  \nlcoef [\vh]_0 \cdot \gradh [\rho]_0} - [w]_1 = 0, 
%						\label{app:qg4}	\\
%	&& \Dt {[\vh]_0} + \nlcoef [\vh]_0 \cdot \gradh [\vh]_0
%		+ \ve k \times [\vh]_1 + \beta y \ve k \times [\vh]_0
%		= - \gradh [P]_1,		\label{app:qg5}
%	\end{eqnarray}
%
and
	\begin{eqnarray}
	&& [w]_1 = -\frac {\Fr^2} {N^2} {\mathscr L} 
			\parenth{\Dz {[P]_0}}, 		\label{app:qg6} \\
	&& [\vh]_1 = \ve k \times {\mathscr L}([\vh]_0) - \beta y [\vh]_0
			+ \ve k \times \gradh[P]_1,     \label{app:qg7}
	\end{eqnarray}
where $\mathscr L$ is the substantial differential operator along
the geostrophic flow $[\vh]_0$: $\mathscr L= \Dt\ + [\vh]_0 \cdot \gradh$.
Eqs.~(\ref{app:qg1}) - (\ref{app:qg7}) 
are to be used in the text in section~\ref{sect:QG}.

As conventional, let $[P]_0 \equiv \psi$. Following the derivations in
standard textbooks (e.g., \cite{McWilliams_book}), we have
%	\begin{eqnarray*}
%	\Dt\ {\gradh^2\psi} + \nlcoef \ve v_g \cdot \gradh (\gradh^2\psi)
%		+ \gradh \cdot [\vh]_1 + \beta v_g = 0.
%	\end{eqnarray*}
% But by the continuity equation and Eq.~(\ref{app:qg6}),
%	\begin{eqnarray*}
%	\gradh\cdot[\vh]_1 
%	&=& -\Dz{[w]_1}
%	= -\Dz\ \cbrace{\frac{\Fr^2}{N^2} 
%	        \bracket{\Dt\ (\Dz\psi) + 
%			 \nlcoef \ve v_g \cdot \gradh(\Dz\psi)  
%			}}	\\
%	&=&
%	  \Dt\ \bracket{\Dz\ \parenth{\frac{\Fr^2}{N^2} \Dz\psi }}
%	  + \nlcoef \ve v_g \cdot \gradh\bracket{\Dz\ 
%			\parenth{\frac{\Fr^2}{N^2} \Dz\psi }}  \\
%	&&\qquad 
%	  + \nlcoef \frac {\Fr^2} {N^2} 
%			\Dz{\ve v_g} \cdot \gradh
%			\parenth{\Dz\psi}
%	\end{eqnarray*}
% By the thermal wind relation, 
%	$\Dz{\ve v_g} \cdot \gradh \parenth{\Dz\psi} = 0,$
% so the last term vanishes, and hence
	\begin{eqnarray}	\label{app:qg8}
        \Dt\ \bracket{\gradh^2\psi + 
		\Dz\ \parenth{\frac{\Fr^2}{N^2}\Dz\psi}}  +
         \nlcoef J\parenth{\psi,\ \bracket{\gradh^2\psi + 
		\Dz\ \parenth{\frac{\Fr^2}{N^2}\Dz\psi}}
	  } 
	 + \beta \Dx\psi = 0,
	\end{eqnarray}
where $J$ is the Jacobian operator. This is the very 
quasi-geostrophic equation for which we derive the multiscale energetics.

%   \begin{figure*} [t]
%   \begin{center}
%   \includegraphics[angle=0,width=0.75\textwidth] {LR05_phase_error.eps}
%   \caption
%         {
%% An example of phase error in the computed canonical transfer. 
%	 The 300-m APE transfer (in $m^2 s^{-3}$) 
%	 in the Iceland-Faeroe frontal 
%  	 region from the large-scale window to the
%         meso-scale window during a meandering event on
%	 August 21, 1993 
%		(adapted from \cite{LR05}). 
%	 (a) The original map with phase error. 
%	 (b) The horizontally filtered map.
%         \protect{\label{fig:phase_error}}}
%   \end{center}
%   \end{figure*}

%%%%%%%%%%%%%%%%%%%%%%%%%%%%%%%%%%%%%%%%%%%%%%%%%%%%%%%%%%%%%%%%%%%%%%%

%	\begin{figure*}	[h]
%	\begin{center}
%	\includegraphics[angle=0, width=0.65\textwidth] {MJO_buoy.eps}
%	\caption{
%	Rate of buoyancy conversion $b^1$.
%	\protect{\label{fig:MJO_buoy}}}
%	\end{center}
%	\end{figure*}

\end{document}